\documentclass[twocolumn,showpacs,pra,amsmath]{revtex4}
\usepackage{epsfig,graphicx,bm}

\def\be{\begin{equation}}
\def\ee{\end{equation}}
\def\e#1{\label{#1}\end{equation}}
\def\bea{\begin{eqnarray}}
\def\eea{\end{eqnarray}}
\def\ea#1{\label{#1}\end{eqnarray}}
\def\rqn#1{(\ref{#1})}
\def\bes#1{\begin{subequations}\label{#1}}
\def\ese{\end{subequations}}

\begin{document}
\title{Two-qubit decoherence mechanisms revealed via quantum process tomography}
%Analysis of quantum process tomography of multipartite systems
%with applications to superconducting qubits
\author{A.\ G.\ Kofman}
\author{A.\ N.\ Korotkov}
\affiliation{Department of Electrical Engineering, University of
California, Riverside, California 92521}
\date{\today}

\begin{abstract}
We analyze the quantum process tomography (QPT) in the presence of
decoherence, focusing on distinguishing local and non-local
decoherence mechanisms for a two-partite system from experimental
QPT data. In particular, we consider the $\sqrt{\rm iSWAP}$ gate
realized with superconducting phase qubits and calculate the QPT
matrix $\chi$ in the presence of several local and non-local
decoherence processes. We determine specific patterns of these
decoherence processes, which can be used for a fast identification
of the main decoherence mechanisms from an experimental
$\chi$-matrix.

    \end{abstract}
    \maketitle

\section{Introduction}

Quantum information processing is presently a focus of  significant
interest, since it shows promise to perform various computational
and communication tasks which are difficult or impossible to perform
by classical means \cite{nie00}.
 A standard scheme of quantum information processing involves a
sequence of unitary operations (gates) on single qubits or pairs of
qubits.
 Due to coupling to environment, the quantum-processor evolution
suffers from decoherence, which introduces errors into quantum
information processing.
    Effects of decoherence on the desired quantum evolution can be
characterized by a variety of methods jointly called Quantum Process
Tomography (QPT), such as the standard QPT \cite{nie00,poy97,chu97},
ancilla-assisted process tomography (AAPT)
\cite{leu03,dar01,dar03,alp03}, and direct characterization of
quantum dynamics \cite{moh06,moh08}.
 The standard QPT is simplest of the above methods in the sense that
it can be performed with initial states being product states and
local measurements of the final states.

In recent years the QPT has been demonstrated experimentally in
optics \cite{alp03,mit03,maz03,obr04,nam05,lan05,kie05,wan07}, NMR
\cite{Childs-01,wei04,kam05}, for ions in traps
\cite{rie06,Monz-09}, and for solid-state qubits
\cite{how06,Martinis-QPT,Katz-uncollapsing,bia07}.
 One-qubit \cite{alp03,maz03,wan07,how06,Martinis-QPT,Katz-uncollapsing}, two-qubit
\cite{mit03,obr04,nam05,lan05,kie05,kam05,rie06,bia07}, and
three-qubit \cite{wei04,Monz-09} systems have been studied.
Experiments on the QPT involving more than one qubit usually use the
standard QPT.
    The QPT experiments with the superconducting phase qubits
\cite{Martinis-QPT,Katz-uncollapsing,bia07}, which are of the most
interest for us here, have been also based on the standard QPT.
 In the present paper we also use the standard QPT.

The QPT provides a very rich (complete) information on the
performance of a quantum circuit. For $N$ qubits the QPT matrix
$\chi$ \cite{nie00} is generally characterized by $16^N$ real
parameters; this number reduces to $16^N-4^N$ parameters if we limit
ourselves by trace-preserving processes.
 Only $4^N-1$ of these parameters correspond to a unitary
evolution, while the rest of them are due to decoherence.
    The problem of converting experimental QPT data into
a characterization of decoherence processes is of significant
theoretical interest
\cite{bou03,Emerson-07,Bendersky-08,Wolf-08,Mohseni-08,moh08a}.
 However, there is still no good
understanding of the relation between the $\chi$-matrix elements and
decoherence parameters, except for the case of one qubit. Recently,
estimation of one-qubit decoherence parameters by a QPT method was
discussed in Ref. \cite{Mohseni-08} for a specific decoherence
model. In practice, decoherence models are often not known in
advance, especially for systems containing more than one qubit. An
additional complication is that often two or more unknown mechanisms
simultaneously cause decoherence. In the present paper, we consider
an approach to identify the decoherence models from the form of the
$\chi$-matrix provided by an experiment. For that we start with
physically reasonable models of decoherence and analyze
corresponding patterns in the $\chi$-matrix.
 If these patterns are
sufficiently specific, then the main decoherence mechanisms can be
identified from an experimental $\chi$-matrix directly, without a
complicated numerical analysis. As a particular example we consider
the $\sqrt{\rm iSWAP}$ gate made of superconducting phase qubits
\cite{mar02,mcd05,ste06} and calculate the $\chi$-matrix in the
presence of several local and non-local decoherence mechanisms,
which can be anticipated for this system. We show that the patterns
of significant elements of the $\chi$-matrix are quite different for
different decoherence mechanisms, that makes their identification
relatively simple, even when two or more decoherence mechanisms
simultaneously affect the system.

The paper is organized as follows.
 In Sec.\ \ref{sec-QPT-general} we review the standard QPT
for a generic system (with some formulas discussed in Appendix
\ref{C'}) and then in Sec.\ \ref{sec-bipartite} we modify this
formalism to make it more convenient for application to a bipartite
system. Section \ref{sec-Markovian} is devoted to a brief discussion
of the Markovian decoherence and calculation of its contribution
into the $\chi$-matrix. In Sec.\ \ref{sec-nonlocal} we introduce
quantitative characteristics of the decoherence non-locality, which
can be obtained from experimental QPT data.
 Section \ref{VII} is the major
part of our paper, in which we analyze the two-qubit $\sqrt{\rm
iSWAP}$ gate made of superconducting phase qubits. We start with the
discussion in Sec.\ \ref{VIIA} of an ideal $\sqrt{\rm iSWAP}$ gate,
then in Sec.\ \ref{VIIB} we discuss several applicable models of
local and non-local decoherence In Sec.\ \ref{VIIC} these models are
used for the calculation of the $\chi$-matrix of the trivial
(identity) two-qubit gate, then in Sec.\ \ref{VIID} the
$\chi$-matrix of the $\sqrt{\rm iSWAP}$ gate is calculated for the
same decoherence models (it happens to have significant similarities
with the identity gate case), and these results are discussed in
Sec.\ \ref{VIIE}.
 In Sec.\ \ref{VIIF} we analyze effects of decoherence on the $\chi$-matrix in the
case of coupled but strongly detuned qubits.
 Section \ref{VIII} is the brief conclusion.

\section{QPT basics}
\label{sec-QPT-basics}

\subsection{QPT for a generic system}
\label{sec-QPT-general}

According to quantum mechanics, a closed system undergoes a unitary
evolution determined by the system Hamiltonian.
 However, usually quantum systems are coupled to environment, i.e., are open.
 The evolution of an open quantum system is described \cite{nie00,dav76} by a
completely positive linear map ${\cal L}$ (a quantum operation): if
initial density matrix of the system and environment at time $t=0$
is a product state, $\rho^0\otimes\rho^E$, and full evolution is
described by Hamiltonian $H_{\rm SE}$, then at time $t$ the reduced
density matrix of the system only is
   \be \rho={\cal L} [\rho^0], \,\,\,
  \rho_{ij}=\sum_{k,l=0}^{d-1}{\cal L}_{ij,kl}\rho^0_{kl},
 \e{19}
where $d$ is the dimension of the Hilbert space of the system, and
the superoperator ${\cal L}$ has elements
%PM.60.4
 \be
{\cal L}_{ij,kl}=\sum_{i',k',l'}\langle ii'|e^{-iH_{\rm
SE}t/\hbar}|kk' \rangle \langle ji'|e^{-iH_{\rm
SE}t/\hbar}|ll'\rangle^* \rho^E_{k'l'},
 \e{20}
with $i,j,k,l$ denoting orthonormal basis states of the system and
$i',k',l'$ denoting the environment basis states.

 Besides the four-index quantity ${\cal L}_{ij,kl}$, it is
convenient to introduce \cite{bou03,Fujiwara-99} the $d^2\!\times\!
d^2$ matrix $\bm{\mathcal L}$ with the same components, but indexed
in a different way:
 \be
\bm{\mathcal L}_{\langle ij\rangle\langle kl\rangle}={\cal
L}_{ij,kl},
 \e{32}
where we use notation
    \be
\langle ij \rangle= d\,i+j,
    \label{notation<>}    \ee
  so that $\langle ij\rangle=
0,1,\dots,d^2-1$ (notice mnemonic rule that the $d$-nary
representation of the number $\langle ij\rangle$ is ``$ij$").
 Now Eq.\ \rqn{19} can be recast as
 \be
{\bm\rho}=\bm{\mathcal L}{\bm\rho^0},
 \e{2}
in which ${\bm\rho}$ is a column vector obtained by placing the rows
of $\rho$ one after another and then transposing the result,
${\bm\rho}_{\langle ij\rangle}=\rho_{ij}$.

The standard QPT  \cite{nie00,chu97} is based on a different but
equivalent description of a quantum operation:
%TO.14.3
 \be
\rho = {\cal L}[\rho^0]=\sum_{m,n=0}^{d^2-1}\chi_{mn}E_m\rho^0
E_n^\dagger,
 \e{1}
where $E_n$ are linearly independent operators (in $d$-dimensional
Hilbert space) and $\chi$ is a $d^2\!\times\! d^2$ Hermitian
positive-semidefinite matrix, which fully characterizes the quantum
operation. A quantum operation should not increase the trace of the
density matrix, that leads to the  condition \cite{nie00,obr04}
%TO.14.3
 \be
\sum_{m,n=0}^{d^2-1}\chi_{mn}E_n^\dagger E_m\le I,
 \e{10}
where $I$ is the ($d$-dimensional) identity operator. (For operators
the inequality $A\le B$ means that $B-A$ is a positive operator.)
For trace-preserving operations Eq.\ \rqn{10} becomes an equality,
while trace-decreasing operations correspond to situations when the
system leaves its Hilbert space or we consider a measurement with a
particular result.

    The QPT matrix $\chi$ can be obtained from experimental data
in two steps, first calculating the matrix $\bm{\mathcal L}$ and
then converting it into the $\chi$-matrix.
    To obtain $\bm{\mathcal L}$ one needs to prepare
$d^2$ linearly independent initial states $\rho^0_n$ (chosen out of
experimental convenience), perform the evolution, and measure the
resulting states $\rho_n$ using the quantum state tomography
\cite{nie00,liu05}. Using Eq.\ \rqn{2}, we can write $R=\bm{\mathcal
L}R_0$, where $R$ and $R_0$ are $d^2\!\times\! d^2$ matrices
constructed from $\rho_n$ and $\rho^0_n$ as
 \be
   R_{\langle ij\rangle n}=(\rho_n)_{ij} , \,\,\,
   (R_0)_{\langle ij\rangle n}=(\rho_n^0)_{ij} ,
 \e{135}
so that the $n$th column of $R$ is ${\bm\rho}_n$, and similarly for
$R_0$. Therefore, the matrix $\bm{\mathcal L}$  can be obtained
\cite{poy97} as
 \be
\bm{\mathcal L}=RR_0^{-1},
 \e{14}
where the existence of $R_0^{-1}$ is ensured by the linear
independence of the states $\rho_n^0$.

Calculation of the $\chi$-matrix from $\bm{\mathcal L}$ is the
easiest when the operators $E_n$ used in the definition (\ref{1})
form the ``by-element'' basis
    \be
    F_{\langle ij\rangle}=|i\rangle\langle j|,
\label{101} \ee
    which we will call the ``elementary basis'' $F_n$.
    This is because Eq.\ \rqn{19} can be rewritten in a form similar
    to (\ref{1}),
 \be
\rho=\sum_{m,n=0}^{d^2-1}J_{mn}F_m\rho^0F_n^\dagger,
 \e{4}
where $d^2\!\times\! d^2$ matrix $J$ contains the same elements as
$\bm{\mathcal L}$, but in a different order:
%TO.10.3
 \be
 J_{\langle ij\rangle\langle kl\rangle}=\bm{\mathcal L}_{\langle
ik\rangle\langle jl\rangle}.
 \e{3}
Therefore, for the elementary basis, $E_n=F_n$, we obtain $\chi=J$,
so that the $\chi$-matrix consists of reordered elements of
$\bm{\mathcal L}$. Explicitly, this reordering is the following: 1)
each row of $\bm{\mathcal L}$ is converted into a $d\!\times\!d$
matrix by sequentially placing the strings of $d$ elements below
each other and 2) these matrices are placed from left to right, with
a new row of matrices starting after each $d$ steps.
  (Another reordering of the matrix elements of $\bm{\mathcal L}$ is
also often used in the literature
\cite{cho75,gil05,jam72,leu03,hav03,sal05}: the operator $C$, which
is related to $J$ as
%TO.51.8
$J_{\langle ij\rangle\langle kl\rangle}=C_{\langle ji\rangle\langle
lk\rangle}$ \cite{kof}. Operators $C$ and $J$ are called Choi or
Jamiolkowski operators. Both $C$ and $J$ are Hermitian and
positive-definite.)

To obtain $\chi$-matrix for a general operator basis $E_n$, let us
construct the $d^2\!\times\! d^2$ matrix ${\bm E}$, whose $n$th
column contains all elements of the $d\!\times\! d$ matrix $E_n$, so
that ${\bm E}_{\langle ij\rangle n}=(E_n)_{ij}$.
 Then
%TO.12.8,13
$E_n=\sum_{m=0}^{d^2-1}F_m {\bm E}_{mn}$, and hence
$F_n=\sum_{m=0}^{d^2-1}E_m ({\bm E}^{-1})_{mn}$,
% \e{5}
where ${\bm E}^{-1}$ exists because of the linear independence of
$E_n$. In this way from Eq.\ \rqn{4} we obtain
%TO.13.7
 \be
\chi={\bm E}^{-1} J ({\bm E}^{-1})^\dagger.
 \e{6}

    This expression simplifies in an important special case of
mutually orthogonal operators $E_n$, which satisfy equation
%TO.15.1
 \be
{\rm Tr}(E_n^\dagger E_m)= d\, \delta_{nm},
 \e{7}
where $\delta_{nm}$ is the Kronecker symbol and the convenient
normalization factor $d$ allows us to include the unity operator
into the set $E_n$ (as well as products of Pauli matrices for
multi-qubit systems). Generalization to a different normalization is
trivial -- see below. In this special case
%TO.31.2
$({\bm E}^\dagger {\bm E})_{nm}= \sum_{i,j=0}^{d-1}{\bm E}_{\langle
ij\rangle n}^*{\bm E}_{\langle ij\rangle m}={\rm Tr}(E_n^\dagger
E_m)=d\,\delta_{nm}$, i.e.,
%TO.31.3
${\bm E}^\dagger {\bm E} =d \, I$ (in other words, ${\bm
E}/\sqrt{d}$ is a unitary matrix), and therefore
% \be
%TO.31.13
%D=E^\dagger/Q
% \e{8}
Eq.\ \rqn{6} becomes
%TO.31.14
 \be
\chi=d^{-2}{\bm E}^\dagger J {\bm E}.
 \e{9}
In the case \rqn{7} the calculation of the trace of the both sides
of Eq.\ \rqn{10} results in the inequality
%TO.31.14
 \be
{\rm Tr}\,\chi\le 1,
 \e{11}
which becomes the equality for a trace-preserving map.

An important example of the orthogonal unitary-operator basis $E_n$
[satisfying Eq.\ (\ref{7})] for a system of $N$ qubits is the
so-called Pauli basis, which consists of tensor products of $N$
operators from the set $\{I,X,Y,Z\}$, where $X,Y,Z$ are the Pauli
operators.
 The modified Pauli basis with $Y\rightarrow-iY$ is also
used in the literature, for example in the QPT analysis for one and
two qubits in Refs.\ \cite{nie00,chu97}.

    Notice that if in Eq.\ (\ref{7}) the normalization factor $d$
is replaced by an arbitrary number $Q$, then in Eq.\ (\ref{9}) the
factor $d^{-2}$ is replaced by $Q^{-2}$ and Eq.\ (\ref{11}) becomes
${\rm Tr}\,\chi\le d/Q$. In particular, $Q=1$ for the orthonormal
basis $F_n$ introduced by Eq.\ (\ref{101}); in this case ${\bm
E}=I$, and therefore Eq.\ \rqn{9} reduces to the previous result
$\chi=J$.

    Several useful formulas for the $\chi$-matrix are discussed
in Appendix \ref{C'}.
    Notice that the QPT calculation procedure discussed above is slightly
different and simpler than in Refs.\ \cite{nie00,chu97}; in
particular, it involves an inversion of a $d^2\! \times d^2$ matrix
[Eq.\ \rqn{14}] instead of a pseudoinverse calculation for a
$d^4\!\times d^4$ matrix.

At the end of this subsection let us briefly discuss the idea of the
AAPT \cite{leu03,dar01,dar03,alp03}, even though we will not use it
in this paper. To perform the AAPT on a $d$-level system ${\cal S}$,
one needs a similar $d$-level ancillary  system ${\cal S'}$. The
compound system is prepared in the maximally entangled state
$|\Phi\rangle= d^{-1/2} \sum_{i=0}^{d-1}|ii\rangle$, then the
quantum operation ${\cal L}$ is applied to the system ${\cal S}$
only, and then the resulting density matrix of the compound system
is measured by the quantum state tomography. It is easy to see that
the resulting density matrix is $({\cal L}\otimes{\cal
I})[|\Phi\rangle\langle\Phi|] =d^{-1}J$, where ${\cal I}$ is the
identity map. In this way the matrix $J$ is obtained directly, and
may later be converted into the $\chi$-matrix, as discussed above.
In principle, other initial states can be also used for the AAPT,
however the maximally entangled state $|\Phi\rangle$ is the optimal
one \cite{dar03}.

\subsection{QPT for a bipartite system}
\label{sec-bipartite}

Now let us consider a bipartite system ${\cal S}$ consisting of
subsystems ${\cal S}_1$ and ${\cal S}_2$ with the Hilbert space
dimensions $d_1$ and $d_2$, respectively.
 Then the dimension of the Hilbert space of ${\cal S}$
is $d=d_1d_2$, and as the basis we can use the states
  \be
|j\rangle=|j_1\rangle|j_2\rangle\equiv|j_1j_2\rangle,
 \e{134}
constructed out of orthonormal basis states in two subsystems. We
enumerate the states using slightly generalized notation
(\ref{notation<>}), so that $j=\langle j_1j_2\rangle = d_2j_1+j_2$.

    As discussed in the previous subsection, the $\chi$-matrix can be
calculated by performing the quantum operation on $d^2$ initial
states $\rho_n^0$. It is often convenient to use the product states,
    $\rho^0_{\langle n_1n_2\rangle}=\rho^{(1)}_{n_1}
\otimes\rho^{(2)}_{n_2}$
    % \e{137}
(where $\langle n_1n_2\rangle =d_2^2 n_1+n_2$) with linearly
independent sets of states for each subsystem.
  In this case the calculation of the matrix $\bm{\mathcal L}$ via Eq.\ (\ref{14})
may be simplified; however, this requires some modification
\cite{kof} of Eq.\ (\ref{14}). The reason is that the matrix $R_0$
does not coincide with the Kronecker product $R_0^{(1)}\otimes
R_0^{(2)}$, as may be naively expected, but requires an additional
permutation of rows. As the result, it is easier to calculate first
the matrix
 $\bm{\mathcal L}'=R[(R_0^{(1)})^{-1}\otimes (R_0^{(2)})^{-1}]$,
    %\e{136}
and then obtain $\bm{\mathcal L}$ by permutation of columns,
    $\bm{\mathcal L}'_{m\langle i_1j_1i_2j_2\rangle}=
\bm{\mathcal L}_{m\langle i_1i_2j_1j_2\rangle}$,
    %\e{138}
where the four-number notation in indices is the natural
generalization of the notation (\ref{notation<>}): $\langle i_1 j_1
i_2 j_2\rangle = i_1 d_1d_2^2 + j_1  d_2^2 + i_2 d_2 +j_2 $ and
$\langle i_1 i_2 j_1 j_2\rangle = i_1 d_1d_2^2 + i_2  d_1d_2 + j_1
d_2 +j_2 $.

    In particular, for a a two-qubit system with the initial states
chosen as products of the states \cite{nie00,poy97,chu97}
$\rho^{(1)}_n=\rho^{(2)}_n= |\psi_n\rangle\langle\psi_n|$ with
$|\psi_0\rangle=|0\rangle,\ |\psi_1\rangle=|1\rangle,\
|\psi_2\rangle= (|0\rangle+|1\rangle)/\sqrt{2}$, and
$|\psi_3\rangle= (|0\rangle+i|1\rangle)/\sqrt{2}$, we obtain
 \be
%TO.114.8
(R_0^{(1)})^{-1}=(R_0^{(2)})^{-1}=\left(\begin{array}{cccc}
 1&-(1+i)/2&(-1+i)/2&0\\
0&-(1+i)/2&(-1+i)/2&1\\
0&1&1&0\\
0&i&-i&0
\end{array}\right).
 \e{142}

As discussed in the previous subsection, the calculation of the
$\chi$-matrix is the easiest when for the operator basis $E_n$ we
choose the elementary basis (\ref{101}). Then $\chi=J$, where $J$ is
given by Eq.\ (\ref{3}).
    However, for a bipartite system it is convenient to use
the product of operator bases for each subsystem,
    \be
    E_{\langle n_1n_2\rangle}=E^{(1)}_{n_1}\otimes E^{(2)}_{n_2},
    \e{102}
and the product of the elementary bases for each subsystem is not
the elementary basis (\ref{101}) because of different enumeration.
Therefore, to simplify formulas for a bipartite system, we have to
somewhat modify the formulas for the generic system.
    In particular, for the product $F^{(1)}_{n_1}\otimes F^{(2)}_{n_2}$ of
the elementary bases (\ref{101}), we get $\chi=\tilde{J}$, where
    %TO.98.8
    $\tilde{J}_{\langle i_1k_1i_2k_2\rangle\langle j_1l_1j_2l_2\rangle}=
\bm{\mathcal L}_{\langle i_1i_2j_1j_2\rangle \langle k_1
k_2l_1l_2\rangle}$.
    %\e{104}
(The relation between $\tilde{J}$ and $J$ is
    $ \tilde{J}_{\langle
i_1k_1i_2k_2\rangle\langle j_1l_1j_2l_2\rangle}= J_{\langle
i_1i_2k_1k_2\rangle \langle j_1j_2l_1l_2\rangle}$.)

For the basis (\ref{102}) which uses arbitrary orthogonal subsystem
bases $E^{(1)}_{n}$ and $E^{(2)}_{n}$, satisfying equations
    \be
 {\rm Tr}(E_n^{(1)\dagger} E_m^{(1)})= d_1\, \delta_{nm}, \,\,\,
  {\rm Tr}(E_n^{(2)\dagger} E_m^{(2)})= d_2\, \delta_{nm},
  \e{16}
   the $\chi$-matrix can be expressed via $\tilde{J}$ as
     \be
\chi=d^{-2}({\bm E}^{(1)\dagger}\otimes {\bm E}^{(2)\dagger})
{\tilde J} ({\bm E}^{(1)}\otimes {\bm E}^{(2)}),
 \e{145}
which is similar to Eq.\ (\ref{9}) [a straightforward application of
Eq.\ (\ref{9}) would not have the desired Kronecker-product form].
   Notice that if subsystem bases satisfy the orthogonality
condition (\ref{16}), then the compound basis (\ref{102}) satisfies
the orthogonality condition (\ref{7}) with the normalization factor
$d=d_1 d_2$. Therefore, Eq.\ (\ref{11}) remains valid, so that for a
trace-preserving operation
    ${\rm Tr}\,\chi=1$.
    % \e{13}

In particular, for a two-qubit system and the Pauli basis we have
 \be
%TO.115.3
d=4,\quad {\bm E}^{(1)}= {\bm E}^{(2)}= \left(\begin{array}{rrrr}
 1&0&0&1\\
0&1&-i&0\\
0&1&i&0\\
1&0&0&-1
\end{array}\right).
 \e{144}

\section{Markovian decoherence}
 \label{sec-Markovian}

\subsection{General formalism}
 \label{sec-Markovian-gen}

An important special case of a general quantum evolution is the
Markovian evolution,
 \be
\dot{\rho}=M[ \rho ],
 \e{31}
where $(M[ \rho ])_{ij}=\sum_{k,l=0}^{d-1}M_{ij,kl} \rho_{kl}$ and
the superoperator $M$ is the generator of a quantum Markov semigroup
\cite{dav76}.
 The four-index representation of $M$ can be converted into a
$d^2 \! \times \! d^2$ matrix $\bm{M}$ with $\bm{M}_{\langle
ij\rangle\langle kl\rangle}= M_{ij,kl}$ [similar to Eq.\ \rqn{32}],
and then
 \be
\bm{\mathcal L}=e^{\bm{M}t}.
 \e{33}

    It is often convenient to separate the evolution generator
$M=L_{\rm coh}+L$ into the coherent part, $L_{\rm
coh}\rho=-(i/\hbar)[H,\rho]$ with $H$ being the Hamiltonian of the
system, and the generator $L$ of the incoherent evolution
(decoherence). In the matrix form we have
 \be
\bm{M}=\bm{L}_{\rm coh}+\bm{L}, \,\,\,
    (\bm{L}_{\rm coh})_{\langle
ij\rangle\langle kl\rangle}=i(H_{lj}\delta_{ik}-H_{ik}\delta_{jl}).
 \e{49} %\e{50}
In the present paper we are interested in effects of decoherence,
and therefore we assume that the Hamiltonian $H$ is known.
 Given the matrix $\bm{\mathcal L}$, which can be measured as
discussed in Sec. \ref{sec-QPT-basics}, the matrix $\bm{M}$ can in
principle be extracted by solving Eq.\ \rqn{33} (though the
extraction procedure involves some subtleties)
\cite{how06,bou03,Wolf-08}.
 Then $\bm{L}$ can be obtained from Eq.\ \rqn{49}.

    Following the QPT description \rqn{1},
it is convenient to introduce a $d^2 \! \times \! d^2$ matrix
$\lambda$ defined by the equation
 \be
L [\rho ]=\sum_{m,n=0}^{d^2-1}\lambda_{mn}E_m\rho E_n^\dagger,
 \e{37}
with the same operator basis $E_n$.
    The matrix $\lambda$ is Hermitian, and for a trace-preserving
map has
 \be
{\rm Tr}\,\lambda=0.
 \e{39}

The matrix $\lambda$ is a counterpart of the $\chi$-matrix and has
many similar properties \cite{kof}. In particular, for a bipartite
system and for a product basis $E_n$ satisfying Eqs.\ (\ref{102})
and (\ref{16}), the matrix $\lambda$ is given by an equation similar
to Eq.\ \rqn{145},
 \be
\lambda=d^{-2}({\bm E}^{(1)\dagger}\otimes {\bm E}^{(2)\dagger})\,
\nu \, ({\bm E}^{(1)}\otimes {\bm E}^{(2)}),
 \e{110}
where
    %TO.98.8
    $ \nu_{\langle i_1k_1i_2k_2\rangle\langle
j_1l_1j_2l_2\rangle}={\bm L}_{\langle i_1i_2j_1j_2\rangle\langle
k_1k_2l_1l_2\rangle}$.
    % \e{82}
Notice that $\lambda=\nu$ for the elementary product basis
$F^{(1)}_{n1}\otimes F^{(2)}_{n2}$.

\subsection{Weak decoherence}
  \label{sec-Markovian-weak}

    The decoherence should be relatively weak for a practical
quantum information processing. In this case (for a sufficiently
short time $t$) one can expand $\bm{\mathcal L}$ up to the first
order in $\bm{L}$ and obtain in the interaction representation
%TO.72.7
 \be
\bm{\mathcal L}^{\rm int}=\bm{\mathcal L}^I+ \int_0^td\tau
e^{-\bm{L}_{\rm coh}\tau} \bm{L} \, e^{\bm{L}_{\rm coh}\tau},
 \e{114}
where $\bm{\mathcal L}^I=I$ is the $d^2$-dimensional identity
matrix, and the interaction representation describes the evolution
of
 $\rho^{\rm int}(t)= e^{iHt/\hbar}\rho(t)\, e^{-iHt/\hbar}$.

    Further simplification is possible for a very short time or when
 the secular approximation \cite{coh92} is applicable \cite{kof}. Then the
time-dependent factors in the integrand in Eq.\ \rqn{114} can be
omitted, yielding $\bm{\mathcal L}^{\rm int}=\bm{\mathcal L}^I+
\bm{L}t$ and
 \be
\chi^{\rm int}=\chi^I+\lambda t,
 \e{91}
where $\chi^I$ is the process matrix for the identity map (see
Appendix \ref{C'}).  Unfortunately, the secular approximation is
usually applicable only when the subsystems (qubits or qudits) are
uncoupled and there are no external fields, so that in the
situations typical for quantum information processing (quantum
gates) the simple equation \rqn{91} is not applicable. Notice that
the conversion between the Schr\"odinger and interaction
representations for the $\chi$-matrix is (see Appendix \ref{C'})
$\chi=V\chi^{\rm int}V^\dagger$, where $V$ is a unitary matrix with
$V_{nm}={\rm Tr}(E_n^\dagger e^{-iHt/\hbar}E_m)/d$ for the
orthogonal basis $E_n$ satisfying Eq.\ (\ref{7}).

\section{Characteristics of nonlocal decoherence}
 \label{sec-nonlocal}

QPT provides a wealth of information: there are $d^4$ independent
real parameters in the matrix $\chi$ (or $d^4-d^2$ for a
trace-preserving quantum operation), and the number of these
parameters increases exponentially with the number of subsystems.
   However, the number of independent parameters for a multipartite
system decreases drastically for local (independent) decoherence of
the subsystems. In this section we discuss local decoherence of a
bipartite system (generalization to a multipartite system \cite{kof}
is rather straightforward).

\subsection{Uncoupled subsystems}

Let us start with assuming uncoupled subsystems, so that unitary
evolution is local. If also decoherence is local, it is easy to show
that for the product basis (\ref{102}) the $\chi$-matrix is the
Kronecker product of the corresponding $\chi$-matrices for the
subsystems,
%TO.88.4
 \be
\chi^{\rm unc}=\chi^{(1)}\otimes\chi^{(2)}.
 \e{74}
In this case the number of independent parameters is
$\tilde{n}_\chi=d_1^4+d_2^4$ (or $\tilde{n}_\chi'=
d_1^4+d_2^4-d_1^2-d_2^2$ in the trace-preserving case), which is
much less than for a general $\chi$-matrix: $n_\chi=d_1^4d_2^4$ (or
$n_\chi'=d_1^4d_2^4-d_1^2d_2^2$).
 There is roughly a square-root decrease of complexity ($N$th root
decrease for an $N$-partite system).
 In particular, for a two-qubit system $\tilde{n}_\chi=32$ and
$\tilde{n}_\chi'=24$ versus $n_\chi=256$ and $n_\chi'=240$.

In an experiment it is generally not known in advance whether
decoherence is local or not. Therefore, a quite important
information can be obtained by checking whether or not a given
$\chi$-matrix has the product form \rqn{74} or, more generally, by
quantifying the accuracy of the product-form approximation.

    Let us define the reduced $\chi$-matrices for subsystems as
    \be
\tilde{\chi}^{(1)}= {\rm Tr}_2 \chi , \,\,\,\, \tilde{\chi}^{(2)}=
{\rm Tr}_1 \chi
    \end{equation}
(in more detail, $\tilde{\chi}^{(1)}_{m_1n_1}=\sum_{m_2=0}^{d_2^2-1}
\chi_{\langle m_1m_2\rangle\langle n_1m_2\rangle}$ and similarly for
$\tilde{\chi}^{(2)}$), and introduce
%TO.86.5
 \be
\tilde{\chi}=\tilde{\chi}^{(1)}\otimes \tilde{\chi}^{(2)}.
 \e{75}
A process matrix $\chi$ is factorizable  if and only if
$\chi=\tilde{\chi}$.
 For $\chi=\tilde{\chi}$  in a trace-preserving case when ${\rm
Tr}\,\chi^{(1)}={\rm Tr}\,\chi^{(2)}={\rm Tr}\,\chi=1$, the matrices
$\chi^{(1)}$ and $\chi^{(2)}$ in Eq.\ \rqn{74} necessarily coincide
with $\tilde{\chi}^{(1)}$ and $\tilde{\chi}^{(2)}$.

 If $\chi\ne\tilde{\chi}$, we can introduce a dimensionless parameter
$\epsilon_{\rm NL}$ characterizing non-locality of the decoherence:
    \be
\epsilon_{\rm NL}={\rm Tr}|\chi-\tilde{\chi}|/{\rm
Tr}|\chi-\chi_{\rm ideal}|,
    \e{24}
where $\chi_{\rm ideal}$ is the process matrix for the ideal
coherent operation, which would occur in the absence of decoherence,
and the absolute value of a matrix $A$ is defined as
$|A|=\sqrt{A^\dagger A}$, so that ${\rm Tr}|A|$ is the so-called
``trace norm'' of $A$.
 Since decoherence yields a deviation of $\chi$ from $\chi_{\rm ideal}$,
the reasoning behind the definition (\ref{24}) is comparison of
matrices $\chi -\chi_{\rm ideal}$ and $\tilde{\chi}-\chi_{\rm
ideal}$, and characterization of their relative difference.
     For $\epsilon_{\rm NL}\ll 1$ the factorization \rqn{75} is still a
good approximation, while for $\epsilon_{\rm NL}\sim1$ the
decoherence is significantly non-local. Notice that the definition
(\ref{24}) is meaningful only in the absence of Hamiltonian coupling
between the subsystems.

\subsection{Coupled subsystems}

In the case of Markovian evolution, the nonlocality of decoherence
can be checked even in the presence of a coupling between the
subsystems.
 We assume that the coupling is included into the (known) Hamiltonian
$H$ and that the generator of the incoherent evolution $L$ [and
hence the matrix $\lambda$ -- see Eq.\ (\ref{37})] can be extracted
from experimental data.
 For the case of local decoherence the generators ${\bm L}^{(1)}$
and ${\bm L}^{(2)}$ of the subsystems decoherence contribute to
${\bm L}$ as \cite{kof}
%TO.44.17
 \bea
{\bm L}_{\langle i_1i_2j_1j_2\rangle\langle k_1k_2l_1l_2\rangle} =
 {\bm L}^{(1)}_{\langle i_1j_1\rangle\langle k_1l_1\rangle}
\delta_{i_2k_2}\delta_{j_2l_2} \,\, &&
    \nonumber\\
+ {\bm L}^{(2)}_{\langle i_2j_2\rangle\langle
k_2l_2\rangle}\delta_{i_1k_1}\delta_{j_1l_1}, &&
 \ea{36}
and there is a simple relation
 \bea
\lambda=\lambda^{(1)}\otimes \chi^{I(2)}
+\chi^{I(1)}\otimes\lambda^{(2)},
 \ea{100}
where $\chi^{I(1)}$ and $\chi^{I(2)}$  are the identity-map process
matrices for the subsystems.

    Similar to the discussion above, we can introduce reduced
matrices $\tilde{\lambda}^{(1)}= {\rm Tr}_2 \lambda$ and
$\tilde{\lambda}^{(2)}= {\rm Tr}_1 \lambda$ and their combination
%TO.105.15
 \be
\tilde{\lambda}= \tilde{\lambda}^{(1)}\otimes \chi^{I(2)} +
\chi^{I(1)}\otimes\tilde{\lambda}^{(2)}.
 \e{46}
Also similarly, it can be shown \cite{kof} that a given matrix
$\lambda$ has the local-decoherence form \rqn{100} if and only if
 $\lambda=\tilde{\lambda}$.
In such a case $\lambda^{(1,2)}= \tilde{\lambda}^{(1,2)}$, assuming
trace-preserving operation with ${\rm Tr}\,\lambda^{(1)}={\rm
Tr}\,\lambda^{(2)}={\rm Tr}\,\lambda=0$. When
$\lambda\ne\tilde{\lambda}$, the nonlocality of decoherence can be
characterized by the dimensionless parameter
 \be
\epsilon_{\rm NL}'={\rm Tr}|\lambda-\tilde{\lambda}|/{\rm
Tr}|\lambda|.
 \e{48}

Note that the nonlocality parameters $\epsilon_{\rm NL}'$ and
$\epsilon_{\rm NL}$ are invariant under a change of the bases
$E_n^{(1,2)}$, which preserves orthogonality [Eq.\ \rqn{16}].
Analysis \cite{kof} shows that $\epsilon_{\rm NL}' \approx
\epsilon_{\rm NL}$ when decoherence is weak, the subsystems are
uncoupled for coherent evolution, and either there is also no
coherent evolution of the subsystems or the secular approximation
holds.

\section{Effects of decoherence mechanisms on two-qubit $\sqrt{\rm iSWAP}$
gate}
 \label{VII}

    Even for only two qubits, the number of decoherence parameters
in the $\chi$-matrix is quite big: in a trace-preserving case we
have $d^4-2d^2+1=225$ parameters. This corresponds to the number of
generally possible decoherence processes. Obviously, interpretation
of experimental $\chi$-matrix data in such a case is quite
difficult. However, instead of considering all general decoherence
processes, it is meaningful to consider only physically reasonable
mechanisms. Then by identifying specific features of these
mechanisms in the $\chi$-matrix and comparing with experimental
data, it is possible to find the magnitudes of various decoherence
processes.

    In this section we consider the $\sqrt{\rm iSWAP}$ gate made of
superconducting phase qubits \cite{mcd05,ste06} and calculate the
$\chi$-matrix assuming several plausible models of decoherence. We
focus on identification of specific features of the $\chi$-matrix,
which may serve as an evidence for a particular mechanism. In
particular, we emphasize distinguishing local and non-local
decoherence mechanisms.

\subsection{The $\sqrt{\rm iSWAP}$ gate}
\label{VIIA}

The qubit states $|0\rangle$ and $|1\rangle$ of a superconducting
phase qubit \cite{mar02} are the ground and first excited states in
the potential well. In this section (except Sec.\ \ref{VIIF}) we
assume that the two qubits are in exact resonance and use the
rotating frame, which zeroes the Hamiltonians of the individual
(uncoupled) qubits. Then the Hamiltonian of the capacitively coupled
qubits in the rotating frame  has the form
\cite{mcd05,kof07-coupled}
%TO.38.5,6
 \be
H=(\hbar S/2)(|01\rangle\langle10|+|10\rangle\langle01|),
 \e{25}
where $S$ is the coupling strength (we assume that $S$ is real). The
Hamiltonian \rqn{25}, which can be recast as $H=(\hbar S/4)(X\otimes
X+Y\otimes Y)$, is a special case of the exchange Hamiltonian, the
so called $XY$ Hamiltonian. It was extensively discussed in relation
to quantum computation. Estimation of the exchange Hamiltonian by
means of the QPT was discussed in Refs. \cite{Mohseni-08,moh08a}.

  The evolution of the two-qubit system is then described by the unitary
operator
 \bea
&U(t)=&e^{-iHt/\hbar}=|00\rangle\langle00|+|11\rangle\langle11|
\nonumber\\
&&+\cos(St/2)(|01\rangle\langle01|+|10\rangle\langle10|)\nonumber\\
&&-i\sin(St/2)(|01\rangle\langle10|+|10\rangle\langle01|).
 \ea{27}
For a non-integer value of $tS/2\pi$ the gate \rqn{27} is an
entangling gate and therefore, together with one-qubit gates, it is
sufficient for quantum computation \cite{bre02}.
 In particular, $U(\pi/S)$ provides the iSWAP gate \cite{sch03},
while $U(\pi/2S)\equiv U_{\sqrt{\rm iSWAP}}$ is the $\sqrt{\rm
iSWAP}$ gate \cite{ima99}. For phase qubits the operation of the
$\sqrt{\rm iSWAP}$ gate has been demonstrated experimentally
\cite{mcd05,ste06}.

 We use the Pauli basis,
 \be
E_{\langle n_1n_2\rangle}=X_{n_1}\otimes X_{n_2},
 \e{65}
where $\{X_0,X_1,X_2,X_3\}=\{I,X,Y,Z\}$, so that
$\{E_0,E_1,\dots,E_{15}\}=\{I\otimes I,I\otimes X,I\otimes
Y,I\otimes Z,X\otimes I,\dots,Z\otimes Z\}$.
 Note that `$n_1 n_2$' is
the base-4 representation of $\langle n_1n_2\rangle$, e.g.,
$E_9=X_2\otimes X_1\equiv Y\otimes X$.
 The operators \rqn{65} satisfy the orthogonality condition \rqn{7}
with $d=4$, so that ${\rm Tr}(E_n^\dagger E_m)=4\delta_{nm}$.
 Any linear (Kraus) operator $K$ in the 2-qubit Hilbert
space can be represented as
 \be
K=\sum_{n=0}^{15}k_nE_n
 \e{68}
with $k_n={\rm Tr}(E_n^\dagger K)/4$.
 Correspondingly, any quantum operation of the form
$\rho=K\rho^0 K^\dagger$ is described in the Pauli basis by the
process matrix (see Appendix \ref{C'})
%TO.25.3,6
 \be
\chi_{mn}=k_m k_n^*.
 \e{29}

\begin{figure}[htb]
\includegraphics[width=7.cm]{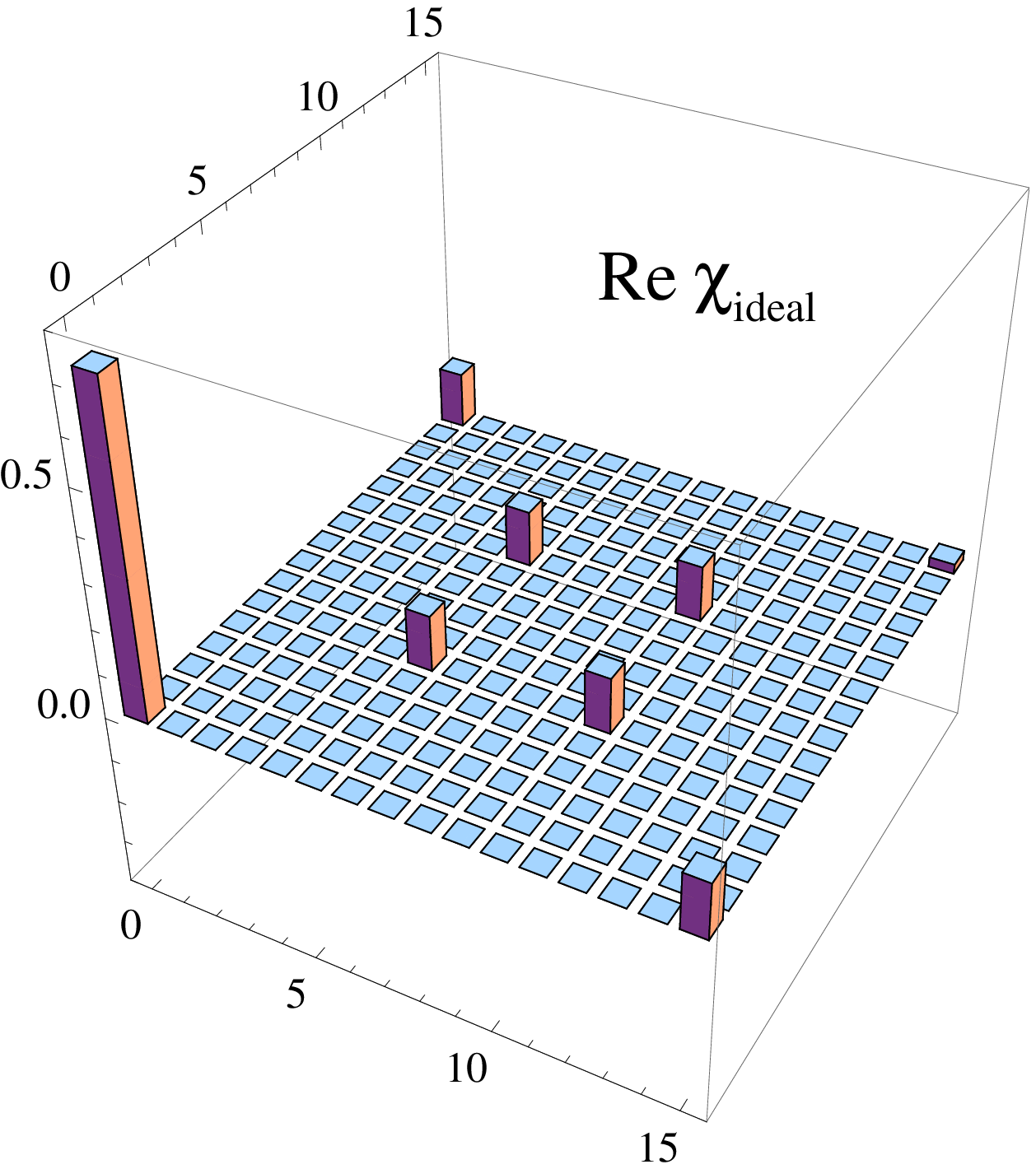}
\includegraphics[width=7.cm]{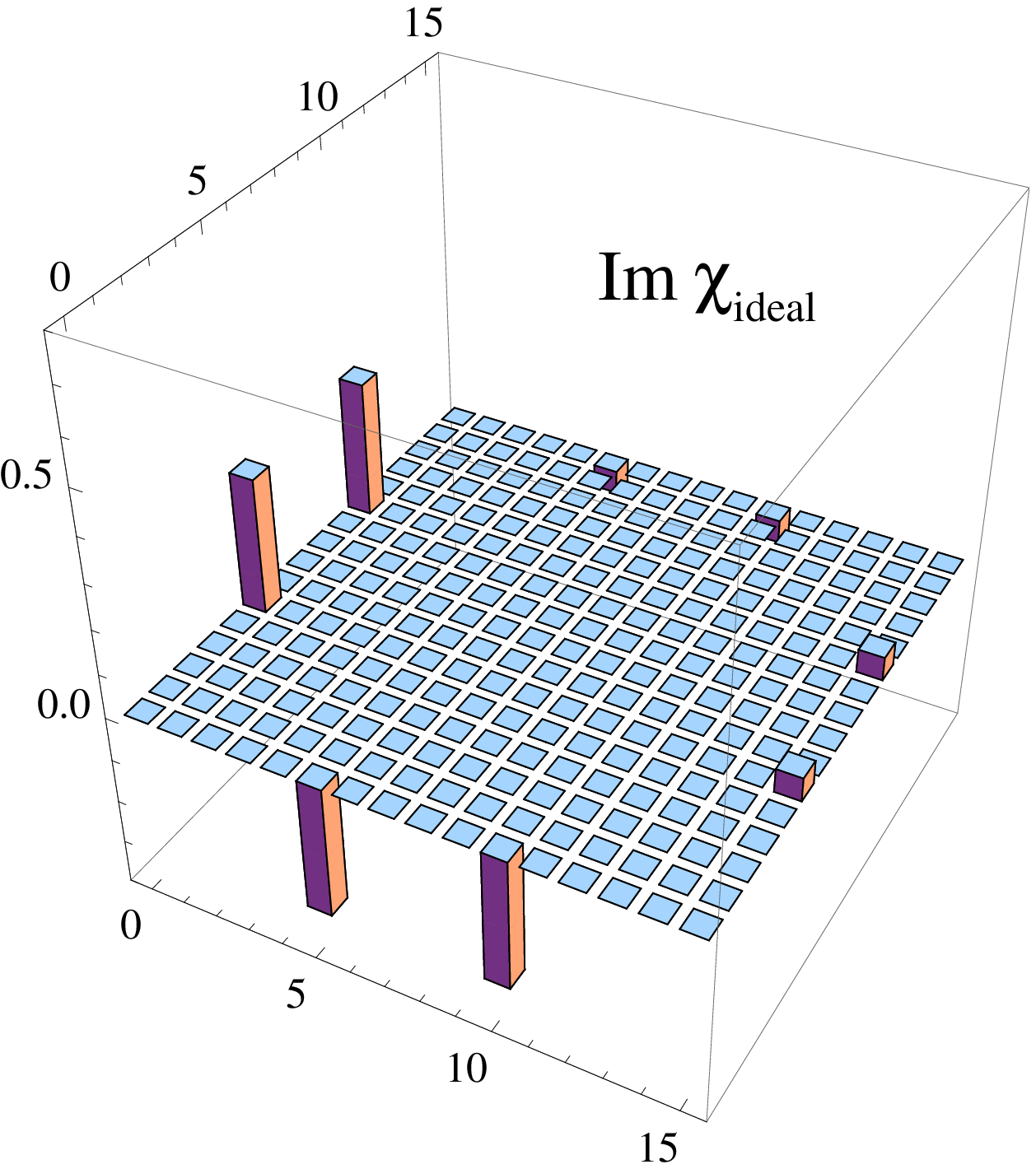}
\caption{The process matrix $\chi_{\rm ideal}$ for the perfect
$\sqrt{\rm iSWAP}$ gate in the Pauli basis.}
 \label{f1}\end{figure}

In the Pauli basis Eqs.\ \rqn{27}--\rqn{68} (with $K=U$) yield
%TO.49.8
 \bea
&U(t)=&\{[1+\cos(St/2)]\, I\otimes I-i\sin(St/2)\, (X\otimes
X\nonumber\\
&&+Y\otimes Y)+[1-\cos(St/2)]\, Z\otimes Z\}/2,
 \ea{149}
and for $t=\pi/2S$ this becomes
%TO.49.5
 \bea
&U_{\sqrt{\rm iSWAP}}=&[(2+\sqrt{2})\, I\otimes
I-i\sqrt{2}\,(X\otimes
X+Y\otimes Y)\nonumber\\
&&+(2-\sqrt{2})\, Z\otimes Z]/4.
 \ea{28}
The process matrices $\chi$ for the gates \rqn{149} and \rqn{28} can
be calculated using Eq.\ \rqn{29}.
 The process matrix $\chi_{\rm ideal}$ for the perfect $\sqrt{\rm iSWAP}$ gate
 is shown in Fig.\ \ref{f1} (since $\chi$ is
Hermitian, the shown elements are symmetric about the main diagonal
in the upper panel and antisymmetric in the lower panel). The
nonzero elements of the matrix $\chi_{\rm ideal}$ are
    \begin{eqnarray}
&& \chi_{00}=(3+2\sqrt{2})/8,\,\,\,\, \chi_{15,15}=(3-2\sqrt{2})/8,
 \nonumber \\
&&
\chi_{55}=\chi_{10,10}=\chi_{5,10}=\chi_{10,5}=\chi_{0,15}=\chi_{15,0}=1/8,
 \nonumber \\
&& \chi_{05}=\chi_{0,10}=-\chi_{50}=-\chi_{10,0}= i(\sqrt{2}+1)/8,
 \nonumber \\
&&
\chi_{15,5}=\chi_{15,10}=-\chi_{5,15}=-\chi_{10,15}=i(\sqrt{2}-1)/8.
\qquad \,\,
\end{eqnarray}

An advantage of using the Pauli basis for the $\chi$-matrix of
$\sqrt{\rm iSWAP}$ is that it results in a relatively small number
of nonzero elements (8 real and 8 imaginary ones) out of the total
number 256.
 For comparison, in the elementary basis
$\{|i_1i_2\rangle\langle j_1j_2|\}$ the operator $U_{\sqrt{\rm
iSWAP}}$ has six terms [see Eq.\ \rqn{27} with $t=\pi/2S$],
resulting in $6^2=36$ nonzero terms in the $\chi$-matrix.
    Notice that conversion of $\chi_{\rm ideal}$ from the Pauli basis
to the modified Pauli basis (with $Y\rightarrow  -iY$) would require
sign change of six elements: $\chi_{10,0}$, $\chi_{10,5}$,
$\chi_{10,15}$, $\chi_{0,10}$, $\chi_{5,10}$, and $\chi_{15,10}$ (in
general this conversion changes 138 out of 256 elements of the
$\chi$-matrix: 18 elements change sign, 60 elements are multiplied
by $i$, and 60 elements are multiplied by $-i$).

 In the presence of decoherence, the $\chi$-matrix
typically acquires additional nonzero elements (in comparison with
$\chi_{\rm ideal}$).
 As shown below, the positions of the most significant extra elements
of $\chi$ may reveal the main mechanisms responsible for
decoherence.

\subsection{Models of decoherence}
\label{VIIB}

In this subsection we consider several physically reasonable
decoherence models for two phase qubits (all Markovian and
trace-preserving), including local decoherence
\cite{sam03,bee03,tyu04,tol05,kof08,kof08a}, correlated dephasing
\cite{ave03,Nor06} and noisy coupling.
    In Refs.\
\cite{sam03,bee03,tyu04,tol05,kof08,kof08a,ave03,Nor06} these models
have been mainly used to analyze two-qubit entanglement and
Bell-inequality violation, while in this paper we focus on their
effect on the $\chi$-matrix of a quantum gate.
 Notice that estimation of one-qubit decoherence parameters by the
QPT was discussed in Ref. \cite{Mohseni-08}.

\subsubsection{Local decoherence}

First, let us consider the model of local decoherence, described by
the Bloch equations \cite{coh92} for each qubit, so that in the
rotating frame the density matrix of a separated qubit $\alpha\
(\alpha=1,2)$ evolves as
%TO.45.1,2
 \bea
&&\dot{\rho}_{11}^{(\alpha)}=-\dot{\rho}_{00}^{(\alpha)}
=-\Gamma_d^{(\alpha)}\rho_{11}^{(\alpha)}
+\Gamma_u^{(\alpha)}\rho_{00}^{(\alpha)},\nonumber\\
&&\dot{\rho}_{10}^{(\alpha)} =-\rho_{10}^{(\alpha)}/T_2^{(\alpha)},
    \ea{4.1}
where $\Gamma_d^{(\alpha)}$ and $\Gamma_u^{(\alpha)}$ are the energy
relaxation rates, so that $T_1^{(\alpha)}= (\Gamma_d^{(\alpha)}+
\Gamma_u^{(\alpha)})^{-1}$ is the energy-relaxation time, and
$T_2^{(\alpha)}$ is the dephasing time (the Bloch equations
correspond to the secular approximation for a non-degenerate
two-level system weakly coupled to a bath; $\Gamma_u^{(\alpha)}=0$
for a zero-temperature bath).

 Comparing Eq.\ \rqn{4.1} with the equation $\dot{\bm\rho}_\alpha
=\bm{L}^{(\alpha)}{\bm\rho}_\alpha$, we obtain the one-qubit
Markovian generators
%TO.45.5
 \be
\bm{L}^{(\alpha)}=\left(\begin{array}{cccc}
 -\Gamma_u^{(\alpha)}&0&0&\Gamma_d^{(\alpha)}\\
0&-1/T_2^{(\alpha)}&0&0\\
0&0&-1/T_2^{(\alpha)}&0\\
\Gamma_u^{(\alpha)}&0&0&-\Gamma_d^{(\alpha)}
\end{array}\right) ,
 \e{3.69}
while the two-qubit generator ${\bm L}_\text{loc}$ of the local
decoherence is then given by Eq.\ \rqn{36}, in which $\langle
ijkl\rangle=8i+4j+2k+l$ (so that `$ijkl$' is the binary
representation of $\langle ijkl\rangle$).

Notice that the model of local decoherence involves two decoherence
mechanisms: energy relaxation and pure dephasing.
 Correspondingly, ${\bm L}_\text{loc}={\bm L}_\text{loc,ER}+
{\bm L}_\text{loc,PD}$. Technically, this splitting corresponds to
representing dephasing rates as sums of two terms,
$1/T_2^{(\alpha)}= (\Gamma_d^{(\alpha)}
+\Gamma_u^{(\alpha)})/2+\Gamma_\alpha$, and then zeroing either
$\Gamma_\alpha$  or $\Gamma_{d,u}^{(\alpha)}$.

\subsubsection{Correlated dephasing}
\label{VB2}

Now let us consider two models of non-local decoherence, starting
with the model of correlated pure dephasing. For a pair of coupled
phase qubits, the correlated dephasing can result from fluctuations
of a common part of the magnetic field biasing qubits. We consider
the system Hamiltonian $H+H_\text{CD}(t)$, in which $H$ is given by
Eq.\ (\ref{25}), while the dephasing contribution  is
%TO.25.3,6
 \bea
&H_\text{CD}(t)=&\hbar\{\delta_1(t)|10\rangle\langle10|
+\delta_2(t)|01\rangle\langle01|\nonumber\\
&&+[\delta_1(t)+\delta_2(t)]|11\rangle\langle11|\},
 \ea{52}
where $\delta_1(t)$ and $\delta_2(t)$ are random but partially
correlated frequency shifts for the two qubits.
 In the derivation of Eq.\ \rqn{52} we neglected noise-induced
transitions between the levels, assuming that the noise intensity at
the qubit frequency practically vanishes.

 Applying the standard method \cite{dav76,red57,Blum}, we obtain the
Markovian master equation for the average density matrix:
 \be
\dot{\rho}=-(i/\hbar)[H,\rho]+L_{\rm CD}[\rho ],
 \e{53}
%TO.70.1-6
 \be
L_{\rm CD}[\rho ]= - \left(\begin{array}{cccc}
 0&\Gamma_2\rho_{01}&\Gamma_1\rho_{02}&\Gamma_+\rho_{03}\\
\Gamma_2\rho_{10}&0&\Gamma_-\rho_{12}&\Gamma_1\rho_{13}\\
\Gamma_1\rho_{20}&\Gamma_-\rho_{21}&0&\Gamma_2\rho_{23}\\
\Gamma_+\rho_{30}&\Gamma_1\rho_{31}&\Gamma_2\rho_{32}&0
\end{array}\right) ,
 \e{51}
where
%TO.69.2-4;70.8
$\Gamma_\pm=\Gamma_1+\Gamma_2\pm\bar{\Gamma}$, $\Gamma_{\alpha}=
\int_{0}^\infty \langle\delta_\alpha(0)\delta_\alpha(t)\rangle \,
dt$ ($\alpha=1,2$), $\bar{\Gamma}= \int_{0}^\infty
\langle\delta_1(0)\delta_2(t) +\delta_2(0)\delta_1(t)\rangle \, dt$,
and we have assumed $\langle\delta_\alpha(t)\rangle=0$.
 The parameter of common dephasing $\bar{\Gamma}$ is zero in the
case of uncorrelated (local) dephasing, while $\bar{\Gamma}=\pm
2\sqrt{\Gamma_1\Gamma_2}$ for full correlation/anticorrelation, so
that the dimensionless correlation parameter is
$\kappa=\bar{\Gamma}/ 2\sqrt{\Gamma_1\Gamma_2}$, $-1\le\kappa\le1$.
In the following subsections we will mainly focus on the case
$\Gamma_1=\Gamma_2\equiv \Gamma_\text{PD}$. Notice that Eq.\
\rqn{51} is written in the computational basis
$|j\rangle=|j_1j_2\rangle$ with $j=\langle j_1j_2\rangle=2j_1+j_2$,
so that $j=0,1,2,3$ correspond to $j_1j_2=00,01,10,11$.
 In deriving Eqs.\ \rqn{53} and
\rqn{51} we have assumed $\Gamma_{\alpha}\tau_c^{\rm CD}\ll1$ and
$S\tau_c^{\rm CD}\ll1$, where $\tau_c^{\rm CD}$ is the correlation
time of the frequency fluctuations.

    In discussion of the QPT it is very easy to get lost with
different bases used in different equations. So we would like to
repeat which bases do we use. In Eq.\ (\ref{51}) [as well as in Eq.\
(\ref{55}) below] we consider a two-qubit density matrix, so this
$4\!\times\!4$ matrix uses the two-qubit basis $\{|00\rangle ,\,
|01\rangle ,\, |10\rangle ,\, |11\rangle \}$. Then this equation is
converted into the equation for the $16\!\times\! 16$ matrix ${\bm
L}$, which uses the basis of 16 elements of the 2-qubit density
matrix. The matrix $\bm{\mathcal L}=e^{(\bm{L}_{coh}+\bm{L})t}$ uses
the same ``by-element'' basis as ${\bm L}$. Finally, the matrix
$\bm{\mathcal L}$ is converted into the $16\!\times\! 16$ matrix
$\chi$, for which we use the basis of product-Pauli operators.
Somewhat differently, in the previous subsection Eq.\ (\ref{4.1})
uses the one-qubit basis $\{|0\rangle ,\, |1\rangle \}$ and Eq.\
(\ref{3.69}) uses still one-qubit but 4-dimensional ``by-element''
basis. Equation (\ref{3.69}) is then converted into the equation for
${\bm L}$, which uses the same 16-dimensional basis as above, and
further procedures coincide. Notice that the bases discussed in this
paragraph have nothing to do with the set of initial states
discussed in Sec.\ \ref{sec-QPT-basics} [e.g., in the paragraph
above Eq.\ (\ref{142})], which would be important in the
experimental procedure.

\subsubsection{Noisy coupling}
\label{VB3}

The second non-local decoherence model we consider is the model of a
noisy coupling.
 In the case of capacitively coupled phase qubits this model
corresponds to a fluctuating coupling capacitance; a more
practically important case is when qubits are coupled via a tunable
Josephson circuit, whose parameters may fluctuate.
    In the Hamiltonian \rqn{25} we substitute  $S$ with $S+s(t)$, assuming
$\langle s(t)\rangle=0$. Then following the same derivation as in
the previous subsection, we obtain the master equation
 \be
\dot{\rho}=-(i/\hbar)[H,\rho]+L_{\rm NC}[\rho ] ,
 \e{54}
where
%TO.71.3-9
 \be
L_{\rm NC}[\rho ]=\Gamma_s\left(\begin{array}{cccc}
 0&-\rho_{01}&-\rho_{02}&0\\
-\rho_{10}&-2\rho_{11}+2\rho_{22}&-2\rho_{12}+2\rho_{21}&-\rho_{13}\\
-\rho_{20}&2\rho_{12}-2\rho_{21}&2\rho_{11}-2\rho_{22}&-\rho_{23}\\
0&-\rho_{31}&-\rho_{32}&0
\end{array}\right)
 \e{55}
and $\Gamma_s=(1/4) \int_{0}^\infty\langle s(0)s(t)\rangle dt$. In
the derivation we have assumed $\Gamma_s\tau_c^{\rm NC}\ll 1$ and
$S\tau_c^{\rm NC}\ll1$, where $\tau_c^{\rm NC}$ is the correlation
time of $s(t)$.

\vspace{0.5cm}

When the discussed above decoherence mechanisms exist concurrently,
the system state obviously evolves as
 \be
\dot{\rho}=-(i/\hbar)[H,\rho]+(L_{\rm loc}+L_{\rm CD}+L_{\rm
NC})[\rho ].
 \e{57}
By fitting experimental data with this model, it is possible to find
the corresponding best-fit decoherence rates quantitatively, and
determine in this way if a particular decoherence mechanism is
important or not. However, this is a rather laborious procedure.
Another way to find out which decoherence mechanisms are important,
is by checking characteristic features in the $\chi$-matrix, unique
for a given mechanism. We will identify such features in the
following subsections.

\subsection{Effects of decoherence on the identity gate}
\label{VIIC}

Before studying the effects of decoherence on the $\chi$-matrix of
the $\sqrt{\rm iSWAP}$ gate (that will be done in the next
subsection), let us consider decoherence for the identity gate,
i.e., for the vanishing two-qubit Hamiltonian. Then $\bm{\mathcal
L}=e^{\bm{L}t}$ with the models for the decoherence generator
$\bm{L}$ discussed above, and $\bm{\mathcal L}$ can be converted
into $\chi$ in the way discussed in Sec.\ \ref{sec-bipartite}.

We are interested in effects of weak decoherence, corresponding to
sufficiently short gate-operation times.
  In this case the process matrix for the identity gate can be
approximated [see Eqs.\ \rqn{91} and (\ref{88})] as
    \be
\chi \approx \chi^I +\lambda t, \,\,\,\, \chi^I_{mn} =
\delta_{m0}\delta_{n0},
    \label{chi-simple-lambda} \end{equation}
where $\lambda$ is determined by Eqs.\ \rqn{110} and \rqn{144} (we
use the Pauli basis). The matrix $\lambda$ is a sum of contributions
from different decoherence mechanisms, which have the following
explicit forms.

For the local energy-relaxation mechanism, the nonzero matrix
elements of $\lambda$ in the Pauli basis are
%TO.102.5-11
 \bea
&&\lambda_{00}=-2(\Gamma_+^{(1)}+\Gamma_+^{(2)}),\nonumber\\
&&\lambda_{11}=\lambda_{22}=\Gamma_+^{(2)},\ \
\lambda_{44}=\lambda_{88}=\Gamma_+^{(1)},\nonumber\\
&&\lambda_{03}=\lambda_{30}=\Gamma_-^{(2)},\ \
\lambda_{0,12}=\lambda_{12,0}=\Gamma_-^{(1)},\nonumber\\
&&\lambda_{21}=-\lambda_{12}=i\Gamma_-^{(2)},\ \
\lambda_{84}=-\lambda_{48}=i\Gamma_-^{(1)},
 \ea{93}
where $\Gamma_\pm^{(\alpha)}=(\Gamma_d^{(\alpha)}\pm
\Gamma_u^{(\alpha)})/4$ [notice a difference with the notation
$\Gamma_\pm$ used in Eq.\ (\ref{51})].
   The contribution from the local pure-dephasing mechanism
is a special case of the correlated dephasing which we discuss next.

For the (correlated) pure dephasing the nonzero matrix elements of
$\lambda$ are
%TO.103.5
 \bea
&&\lambda_{00}=-(\Gamma_1+\Gamma_2)/2,\nonumber\\
&&\lambda_{33}=\Gamma_2/2,\ \ \lambda_{12,12}=\Gamma_1/2,\nonumber\\
&&\lambda_{3,12}=\lambda_{12,3}=
-\lambda_{0,15}=-\lambda_{15,0}=\bar{\Gamma}/4.
 \ea{94}
The absence of the correlation, $\bar{\Gamma}=0$, corresponds to the
local pure dephasing; in this case the third line in Eq.\ \rqn{94}
vanishes.

For the noisy coupling the nonzero elements of $\lambda$ are
%TO.109.1
 \bea
&\lambda_{00}=-\Gamma_s,\ \ \lambda_{55}&=\lambda_{10,10}
=\lambda_{5,10}=\lambda_{10,5}\nonumber\\
&&= \lambda_{0,15}=\lambda_{15,0}=\Gamma_s/2.
 \ea{95}

All nonzero elements of $\chi$ except for $\chi_{00}$, are induced
by decoherence. Because of the first-order approximation
(\ref{chi-simple-lambda}), the most significant additional elements
of $\chi$ are related approximately linearly to the nonzero elements
of $\lambda$ (the second-order elements of $\chi$ should be
significantly smaller for a weak dephasing).
    Now, a very important observation is that
the positions of the nonzero elements of $\lambda$ (excluding
$\lambda_{00}$) in Eqs.\ \rqn{93}--\rqn{95} are different for
different decoherence models, except for $\lambda_{0,15}$ and
$\lambda_{15,0}$ appearing in both Eqs.\ \rqn{94} and \rqn{95}.
 Therefore, in the case of weak decoherence {\it one can identify the
considered decoherence mechanisms simply by the positions of the
most significant (first-order) elements of} $\chi$.
    Another important observation is that effects of different
decoherence parameters on elements of $\lambda$ are easily
distinguishable.
    In particular, for the local decoherence model [Eqs.\ \rqn{93} and \rqn{94}
with $\bar{\Gamma}=0$], the decoherence in the first and second
qubits is completely separated, affecting different elements of
$\lambda$. Similarly, for each qubit the pure dephasing is separated
from energy relaxation by affecting different elements of $\lambda$,
and the temperature for each qubit can be extracted from the ratio
$\Gamma_-^{(\alpha)}/\Gamma_+^{(\alpha)}$, which is equal to the
ratio of corresponding elements of $\lambda$ in Eq.\ \rqn{93}. The
correlation factor in the pure dephasing model can be extracted via
the relative height of elements in the second and third lines of
Eq.\ \rqn{94} (two positive elements in the third line should be
used, since the negative elements are also involved in the noisy
coupling model).
    The clear separation of effects allows us to estimate the relative
values of the decoherence parameters simply by the relative values
of the corresponding elements in the $\chi$-matrix.

    Such a simple analysis is possible, to a large extent, because we
use the Pauli basis. The matrix $\lambda$ in the Pauli basis has a
relatively small number of non-zero elements. The first row in Table
\ref{t1} shows this number for the considered decoherence models.
For comparison, in the elementary operator basis
$\{|i_1i_2\rangle\langle j_1j_2|\}$ we have $\lambda=\nu$  [see Eq.\
(\ref{110})], and then the significantly larger number of non-zero
elements is given by the second row in Table \ref{t1} [the number of
nonzero elements of $\nu$ equals that of $\bm{L}$].
     Since Eq.\ (\ref{chi-simple-lambda}) is only an approximation,
the number of non-zero elements of the matrix $\chi$ is larger than
that for $\lambda$; it is shown in the third row of Table \ref{t1}
for the Pauli basis.  For the elementary basis [then $\chi
=\tilde{J}$, see Eq.\ (\ref{145})] this number is shown in the
fourth row and is typically significantly larger (except for the
model of energy relaxation). This illustrates convenience of the
Pauli basis.

%TO.116.4-11
\begin{table}[tb]
\begin{tabular}{c|cccccccc}
 &ER$_{T>0}$& ER$_{T=0}$& LPD& LD$_{T>0}$& LD$_{T=0}$& CD&
CCD&\ NC\\
\hline
 $\lambda$&13&13&3&15&15&7&7&7\\
  $\nu$&32&23&12&32&23&12&10&16\\
   $\chi$&64&64&4&64&64&8&8&8 \\
 $\tilde{J}$&36&25&16&36&25&16&16&20
 \end{tabular}
\caption{The number of nonzero elements in the matrices $\lambda$,
$\nu$, $\chi$, and $\tilde{J}$ for several decoherence models:
energy relaxation (ER) with arbitrary or zero temperature $T$, local
pure dephasing (LPD), local decoherence (LD) which includes both
pure dephasing and energy relaxation (with $T>0$ or $T=0$),
correlated pure dephasing (CD) with $0<|\kappa| <1$, completely
correlated/anticorrelated pure dephasing (CCD) with $\kappa=\pm1$,
and noisy coupling (NC). We use the Pauli basis for the matrices
$\lambda$ and $\chi$, while in the elementary basis they would be
equal to the matrices $\nu$ and $\tilde{J}$, correspondingly. The
identity gate is assumed for matrices $\chi$ and $\tilde{J}$.
 }
 \label{t1}\end{table}

 \vspace{0.2cm}

   The nonlocality parameters  $\epsilon_{\rm NL}$ and  $\epsilon_{\rm NL}'$
introduced by Eqs.\ \rqn{24} and \rqn{48} approximately coincide in
the weak-decoherence case (\ref{chi-simple-lambda}); $\epsilon_{\rm
NL}'$ is time-independent, while $\epsilon_{\rm NL}$ slowly changes
with time.
    Calculation of $\epsilon_{\rm NL}'$ gives the following results
for the considered decoherence models. For the local decoherence
involving both energy relaxation and pure dephasing, we obtain
%TO.102.12
$\epsilon_{\rm NL}'=0$, as should be expected. For the model of
correlated pure dephasing with $\Gamma_1=\Gamma_2$, the nonlocality
parameter $\epsilon_{\rm NL}'=2|\kappa |/(1+\sqrt{1+\kappa^2})$
depends on the correlation factor $\kappa$, so that $\epsilon_{\rm
NL}'\approx|\kappa|$ for $|\kappa|\ll1$ and
%TO.103.8,11
$\epsilon_{\rm NL}'=2\sqrt{2}-2\approx 0.83$ for $\kappa=\pm 1$.
 Finally, for a noisy coupling
%TO.109.2
$\epsilon_{\rm NL}'=2\sqrt{2}-1\approx 1.83$. As expected,
$\epsilon_{\rm NL}'$ is of the order of 1 for a strongly nonlocal
decoherence.

\subsection{Effects of decoherence on the $\sqrt{\rm iSWAP}$ gate}
\label{VIID}

Now let us consider the effects of decoherence on the $\chi$-matrix
of the $\sqrt{\rm iSWAP}$ gate. We calculate $\chi$ via Eq.\
\rqn{145} from the evolution equation $\bm{\mathcal
L}=e^{(\bm{L}_{\rm coh}+\bm{L})\pi/2S}$, where $\bm{L}_{\rm coh}$ is
given by Eqs.\ \rqn{49} and \rqn{25}, and $16\!\times\! 16$ matrix
${\bm L}$ depends on the decoherence model (Sec.\ \ref{VIIB}).

   In the important case of weak decoherence the first-order
approximation (\ref{114}) leads to the linear relation between the
decoherence contribution $\chi - \chi_{\rm ideal}$ and the
decoherence generator $\bm{L}$ (the evolution time is fixed).
Therefore, since the decoherence generators for various mechanisms
simply add up [see Eq.\ \rqn{57}], their contributions into the
$\chi$ matrix are approximately additive for a weak decoherence,
 \be
\chi\approx\chi_{\rm ideal}+\delta\chi_{\rm loc}+\delta\chi_{\rm CD}
+\delta\chi_{\rm NC},
 \e{56}
and we can consider them separately.

Figures \ref{f2}--\ref{f4} discussed below show the numerical
results for the $\chi$-matrix of the $\sqrt{\rm iSWAP}$ gate in the
presence of the decoherence mechanisms considered in Sec.\
\ref{VIIB}.
 A comparison of Fig.\ \ref{f1} with Figs. \ref{f2}--\ref{f4} shows
that the effect of decoherence on the $\chi$-matrix is to modify the
values of the nonzero elements of $\chi_{\rm ideal}$ and generally
to add extra nonzero elements.
 Below we identify patterns of extra elements specific for each
considered decoherence mechanism, which allow for a fast, though
tentative, identification of these mechanisms.
 For this purpose it is sufficient to consider only the largest
extra elements (most significant out of the first-order in
decoherence elements). An important observation (see below) is that
for the considered decoherence models the positions of the largest
extra elements of $\chi$ coincide with the positions of elements of
$\lambda$ discussed in the previous subsection.
 This makes the analysis of the $\chi$-matrix for the $\sqrt{\rm iSWAP}$
gate rather similar to the analysis in the absence of unitary
evolution (Sec.\ \ref{VIIC}), except for the noisy-coupling model
for which there are no extra elements in $\chi$.
 Consider now the effects of decoherence models in more detail.

\subsubsection{Local decoherence}

We focus on parameter values typical for experiments with the
superconducting phase qubits, and therefore assume zero-temperature
case (which is a very good approximation for the experiments
\cite{Martinis-QPT,Katz-uncollapsing,bia07,mar02,mcd05,ste06}). We
also assume the same local-decoherence parameters for both qubits,
so that $\Gamma_d^{(1)}=\Gamma_d^{(2)}=1/T_1$,
$T_2^{(1)}=T_2^{(2)}=T_2$, and $\Gamma_u^{(1)}=\Gamma_u^{(2)}=0$.

\begin{figure}[htb]
\includegraphics[width=7.cm]{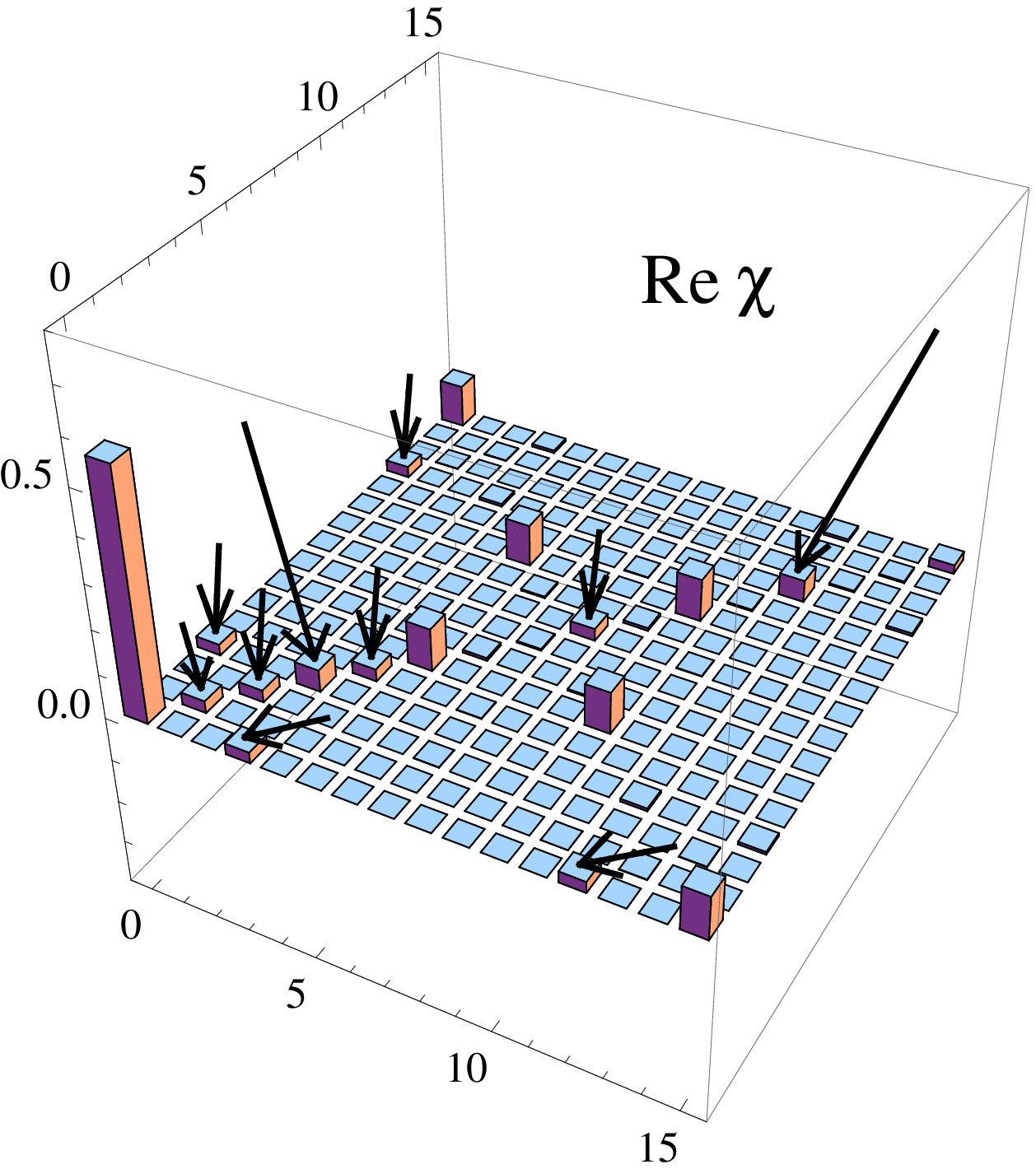}
\includegraphics[width=7.cm]{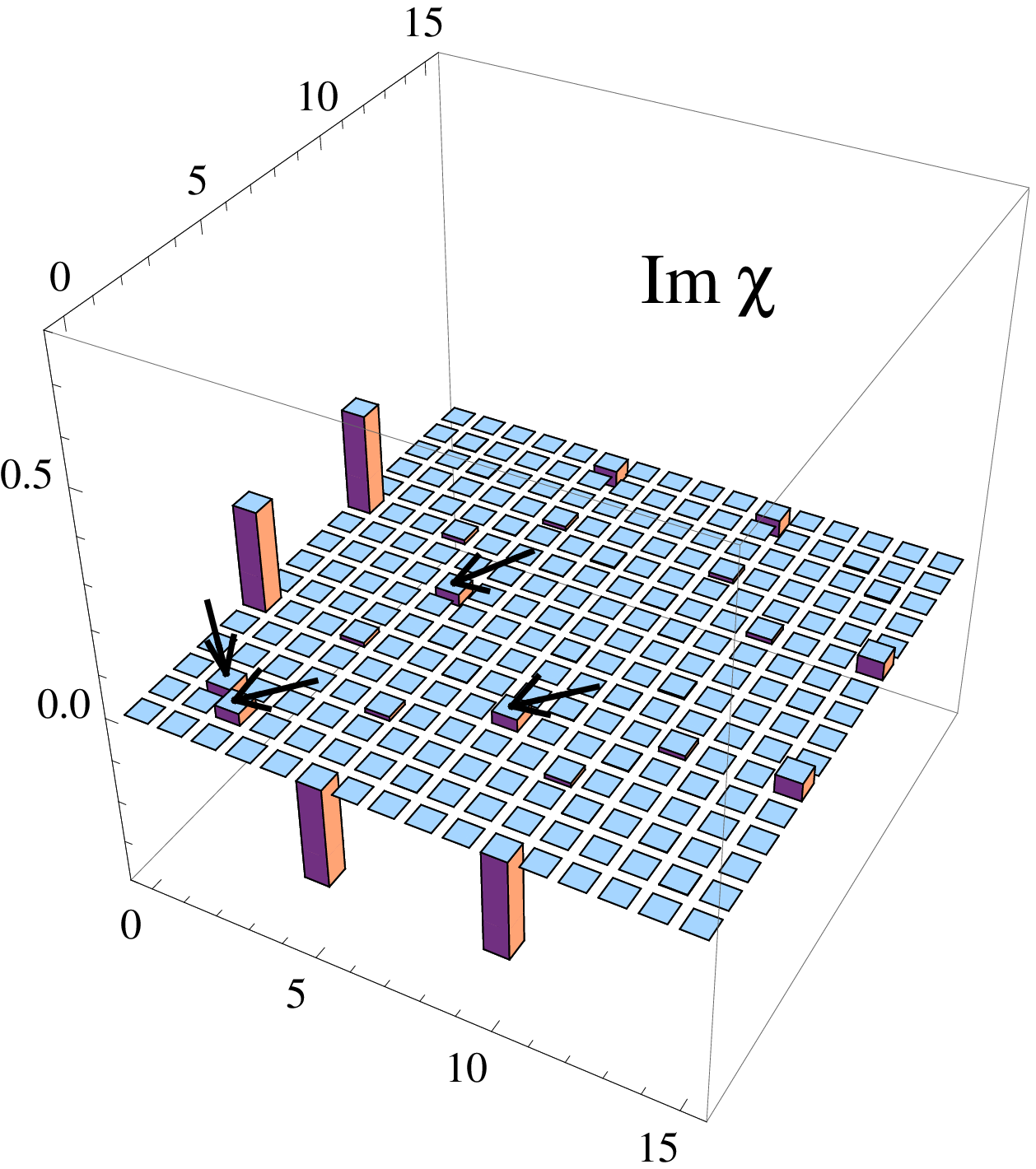}
\caption{The process matrix $\chi$ (in the Pauli basis) for the
$\sqrt{\rm iSWAP}$ gate in the presence of local decoherence  for
$S/2\pi=20$ MHz, $T_1=90$ ns, and $T_2=60$ ns.
 The matrix elements marked by the long arrows,  $\chi_{33}=\chi_{12,12}$,
are the features of the pure dephasing, while the elements marked by
the short arrows ($\chi_{11}= \chi_{22}=\chi_{44}=\chi_{88}$ and
$\chi_{03}=\chi_{30}=\chi_{0,12}=\chi_{12,0}$ in Re$\,\chi$ and
$\chi_{21}=-\chi_{12}=\chi_{84}=-\chi_{48}$ in Im$\,\chi$) are the
features of the energy relaxation.}
 \label{f2}\end{figure}

Figure \ref{f2} shows the process matrix of the $\sqrt{\rm iSWAP}$
gate in the presence of local decoherence. For this example we have
chosen the coupling $S/2\pi=20$ MHz (which is in between the
coupling values of experiments \cite{mcd05} and \cite{ste06}) and
the decoherence parameters $T_1$=90 ns and $T_2$=60 ns, which are
also more or less typical for the superconducting phase qubits (much
longer relaxation times have been achieved recently
\cite{Martinis-Nature08,Martinis-QIP}).

The local decoherence includes two mechanisms: energy relaxation
(with the rate $1/T_1$) and pure dephasing (with the rate
$\Gamma_\text{PD} =1/T_2-1/2T_1$). As follows from the results
presented below, the relative strength of these two mechanisms can
be easily estimated by inspection of the extra elements of $\chi$
(compared to $\chi_{\rm ideal}$) in Fig.\ \ref{f2}. The elements
marked by the long arrows are due to pure dephasing, while the
elements marked by the short arrows are due to the energy
relaxation. By comparing the hight of the elements of $\chi$ marked
by the long and short arrows, one can crudely estimate the relative
strength of these two mechanisms.

 The largest extra elements for the energy-relaxation model (short arrows
in Fig.\ \ref{f2}) in the first order in $1/T_1$ are the following:
 \bea
%TO.95.6-8;101.1
&&\chi_{11}=\chi_{22} =\chi_{44} =\chi_{88}
=\frac{\pi+2\sqrt{2}}{16ST_1}
\approx \frac{0.37}{ST_1},\nonumber\\
&&\chi_{03}=\chi_{0,12} =\chi_{30} =\chi_{12,0}=
\frac{\pi(2+\sqrt{2})}{32ST_1} \approx \frac{0.34}{ST_1},\nonumber\\
&&\chi_{21}=\chi_{84} =-\chi_{12} =-\chi_{48}=
i\frac{\pi+2\sqrt{2}}{16ST_1} \approx i\frac{0.37}{ST_1}.\quad\ \ \
\
 \ea{61}
They are at the same positions as the elements of $\lambda$, Eq.\
\rqn{93} (the remaining element $\lambda_{00}$ is at the location of
the main $\sqrt{\rm iSWAP}$ peak, while we consider only extra
elements of $\chi$-matrix).

    We emphasize that the $\chi$-matrix also contains many other
first-order in $1/T_1$ elements (in contrast to the unity-gate case
considered in the previous subsection); however, they happen to have
relatively small absolute values. The largest of them are imaginary:
$\chi_{46}=\chi_{4,11}=\chi_{13,6}= \chi_{13,11}=
i\pi/(16\sqrt{2}ST_1) \approx 0.14i/ST_1$; this gives 8 elements
together with the corresponding Hermitian conjugated elements. There
are also 4 elements of magnitude $0.06/ST_1$ and 8 elements with the
absolute value $0.02/ST_1$.

 The rest of the extra elements of $\chi$ are of a higher order in
$1/ST_1$, and therefore much smaller than the first-order elements
for $ST_1\gg1$. For instance, $\chi_{33}=\chi_{12,12} =\chi_{3,12}
=\chi_{12,3} \approx (\pi^2/64)(ST_1)^{-2}$ (we show these elements
explicitly because they are located at the same positions as the
most significant elements for the model of pure dephasing discussed
below).

    For the model of local pure dephasing the largest (not all)
first-order in $\Gamma_\text{PD}$ elements are
    \be
\chi_{33}=\chi_{12,12} \approx \frac{(3\pi+2)\,
\Gamma_\text{PD}}{16S} \approx \frac{0.71\, \Gamma_\text{PD}}{S},
    \label{iswap-LPD} \end{equation}
and they are again at the same positions as the elements of
$\lambda$ in Eq.\ \rqn{94}. These elements are shown by the long
arrows in Fig.\ \ref{f2}. Since for Fig.\ \ref{f2} we assumed
$\Gamma_\text{PD}=1/T_1$, the height of these elements is comparable
to (approximately twice larger than) the height of the main extra
elements due to the energy relaxation. The other (much smaller)
first-order in $\Gamma_\text{PD}$ elements are discussed below,
combined with the more general case of correlated pure dephasing,
which we consider next.

\subsubsection{Correlated dephasing}

Let us consider the effects of correlated pure dephasing for
$\Gamma_1=\Gamma_2\equiv\Gamma_\text{PD}$ and arbitrary correlation
factor $\kappa =\bar{\Gamma}/2\Gamma_\text{PD}$.
 Now $\chi$ generally contains eight extra elements,
all of them real. These eight elements are also present in the first
order in $\Gamma_\text{PD}/S$; however, only four of them are
relatively large:
 \bes{62t}
 \bea
%TO.96.8-10
&\chi_{33}=\chi_{12,12}
&\approx \frac{[3\pi+2+(\pi-2)\kappa]\, \Gamma_\text{PD}}{16S}\nonumber\\
 &&\approx (0.71+0.07\kappa)\,\Gamma_\text{PD}/S,\label{66}\\
&\chi_{3,12}=\chi_{12,3} &\approx
\frac{[\pi-2+(3\pi+2)\kappa]\,\Gamma_\text{PD}}{16S}\nonumber\\
 &&\approx (0.07+0.71\kappa)\,\Gamma_\text{PD}/S,
 \ea{62}
 \ese
while the other four elements are much smaller,
$\chi_{66}=\chi_{99}=-\chi_{69}=
-\chi_{96}\approx(\pi-2)(1-\kappa)\Gamma_\text{PD}/(16S) \approx
0.07(1-\kappa)\Gamma_\text{PD}/S$.
 Notice that the larger elements \rqn{62t} are again at the positions
of the elements of $\lambda$ in Eq.\ \rqn{94}, and these positions
are all different from those for the energy relaxation. For a weak
correlation,  $\kappa\ll 1$, the elements $\chi_{3,12}$ and
$\chi_{12,3}$ [see Eq.\ \rqn{62}] become small, recovering the
result (\ref{iswap-LPD}) for the local dephasing.

\begin{figure}[htb]
\includegraphics[width=7.cm]{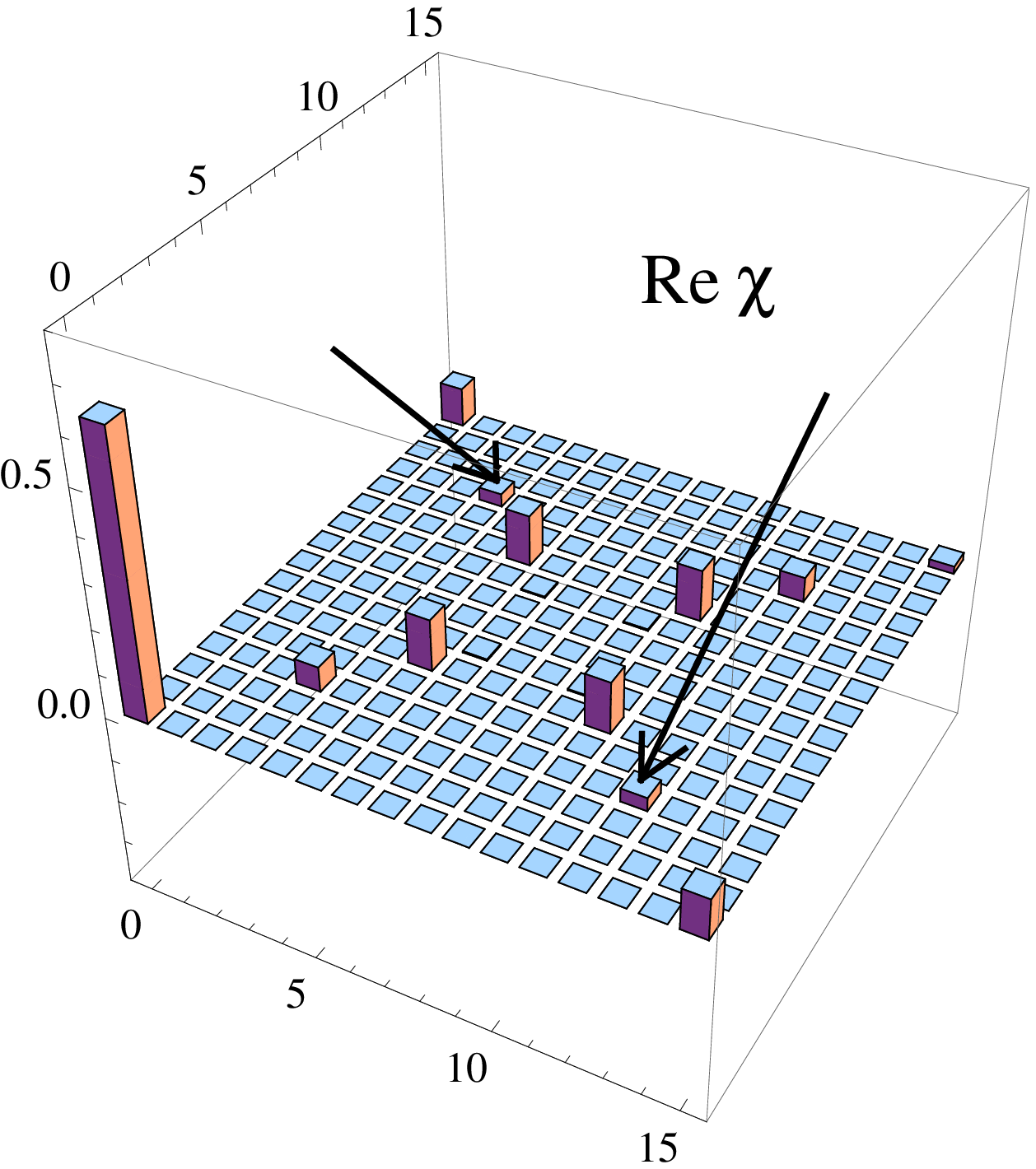}
\includegraphics[width=7.cm]{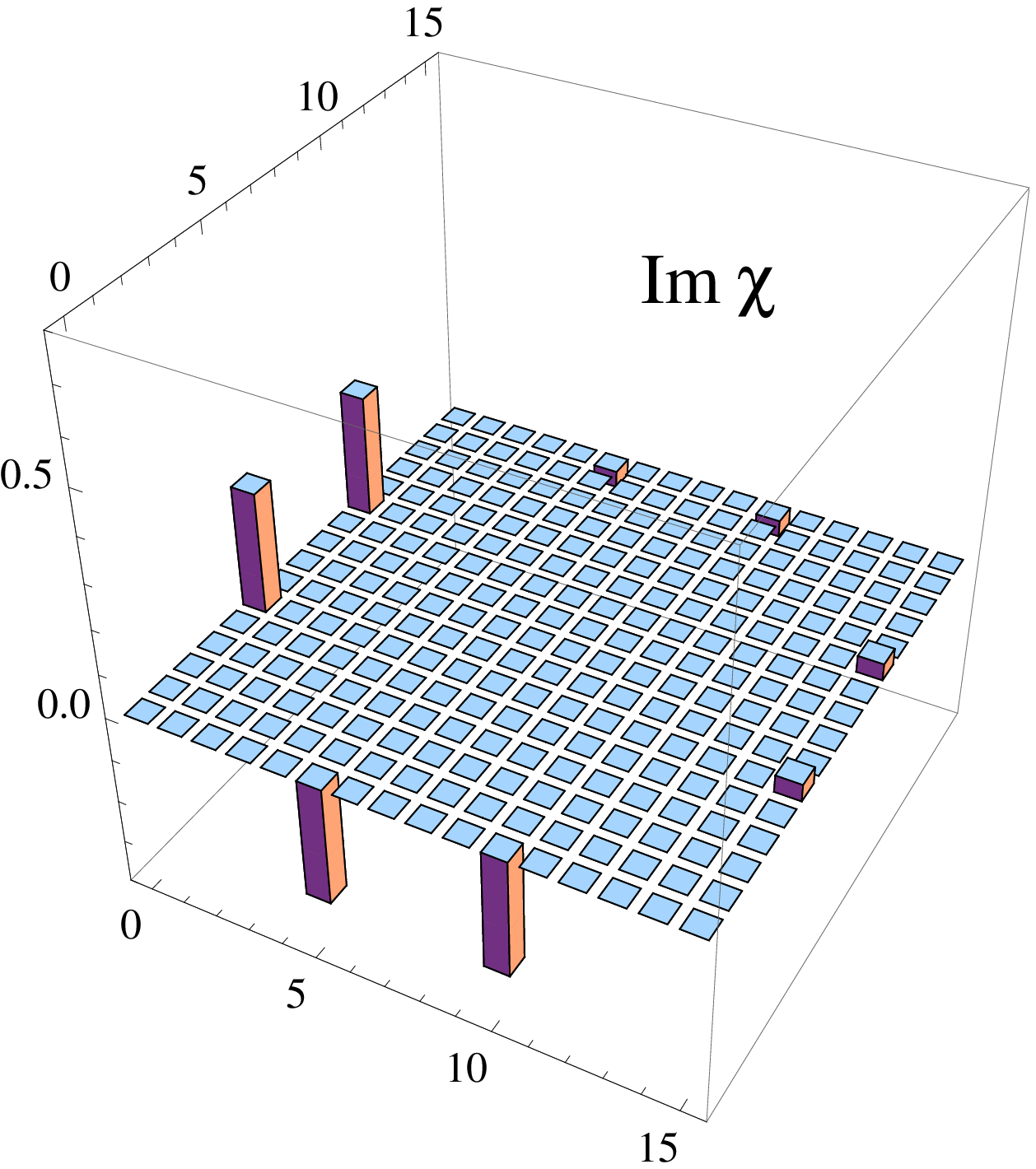}
\caption{The $\chi$-matrix (in the Pauli basis) for the $\sqrt{\rm
iSWAP}$ gate in the presence of correlated pure dephasing for
$S/2\pi=20$ MHz, $\Gamma_\text{PD}=(90$ ns)$^{-1}$, and
$\kappa=0.5$.
 Significant matrix elements shown by the arrows, $\chi_{3,12}=\chi_{12,3}$,
 indicate significant
correlation of the two-qubit dephasing, $|\kappa|\sim1$.}
 \label{f3}\end{figure}

Figure \ref{f3} shows the $\chi$-matrix of the $\sqrt{\rm iSWAP}$
gate affected by the partially correlated dephasing, $\kappa =0.5$.
 The two diagonal extra elements, $\chi_{33}$ and $\chi_{12,12}$,
are at the positions shown by the long arrows in Fig.\ \ref{f2}, and
their values are almost independent of $\kappa$. In contrast, the
off-diagonal elements $\chi_{3,12}$ and $\chi_{12,3}$, marked by the
arrows in Fig.\ \ref{f3}, strongly depend on the correlation
$\kappa$, so that their magnitudes are comparable to the values of
$\chi_{33}$ and $\chi_{12,12}$ only for a significant correlation,
$|\kappa|\sim1$ [see Eq.\ \rqn{62}]. This clearly suggests the way
to check decoherence due to fluctuating common magnetic field in an
experiment with phase qubits.

The $\chi$-matrix has especially simple form in the case of the
completely correlated dephasing, $\kappa=1$, so that $\Gamma_-=0$,
and $\Gamma_+=4\Gamma_\text{PD}$ in Eq.\ (\ref{51}). Then the exact
solution gives the following nonzero elements of $\chi$.
 The elements which are at the positions of the nonzero
elements of $\chi_{\rm ideal}$ become
%TO.77.2,3
 \be
\chi_{mn} =\frac{1}{8}\left(\begin{array}{cccc}
 f_+&ig_+&ig_+&\gamma_d^4\\
-ig_+&1&1&-ig_-\\
-ig_+&1&1&-ig_-\\
\gamma_d^4&ig_-&ig_-&f_-
\end{array}\right) ,
 \e{58}
where $m,n\!=\! 0,5,10,15$, $f_\pm= 2+\gamma_d^4\pm 2\sqrt{2}
\gamma_d,\ g_\pm=\sqrt{2}\gamma_d\pm1$, and $\gamma_d=
e^{-\pi\Gamma_\text{PD}/2S}$.
 The extra nonzero matrix elements are
 \be
\chi_{33}=\chi_{3,12}=\chi_{12,3}=\chi_{12,12}=(1-e^{-2\pi\Gamma_\text{PD}/S})/8,
 \e{59}
so that all of them are equal [as for the first-order result
(\ref{62t}) with $\kappa =1$]. Notice that for a partially
correlated dephasing with $0<\kappa <1$ the exact solution gives
$\chi_{33}=\chi_{12,12} >\chi_{3,12}=\chi_{12,3}>0$ (as in  Fig.\
\ref{f3}).

\subsubsection{Noisy coupling}

\begin{figure}[htb]
\includegraphics[width=7.cm]{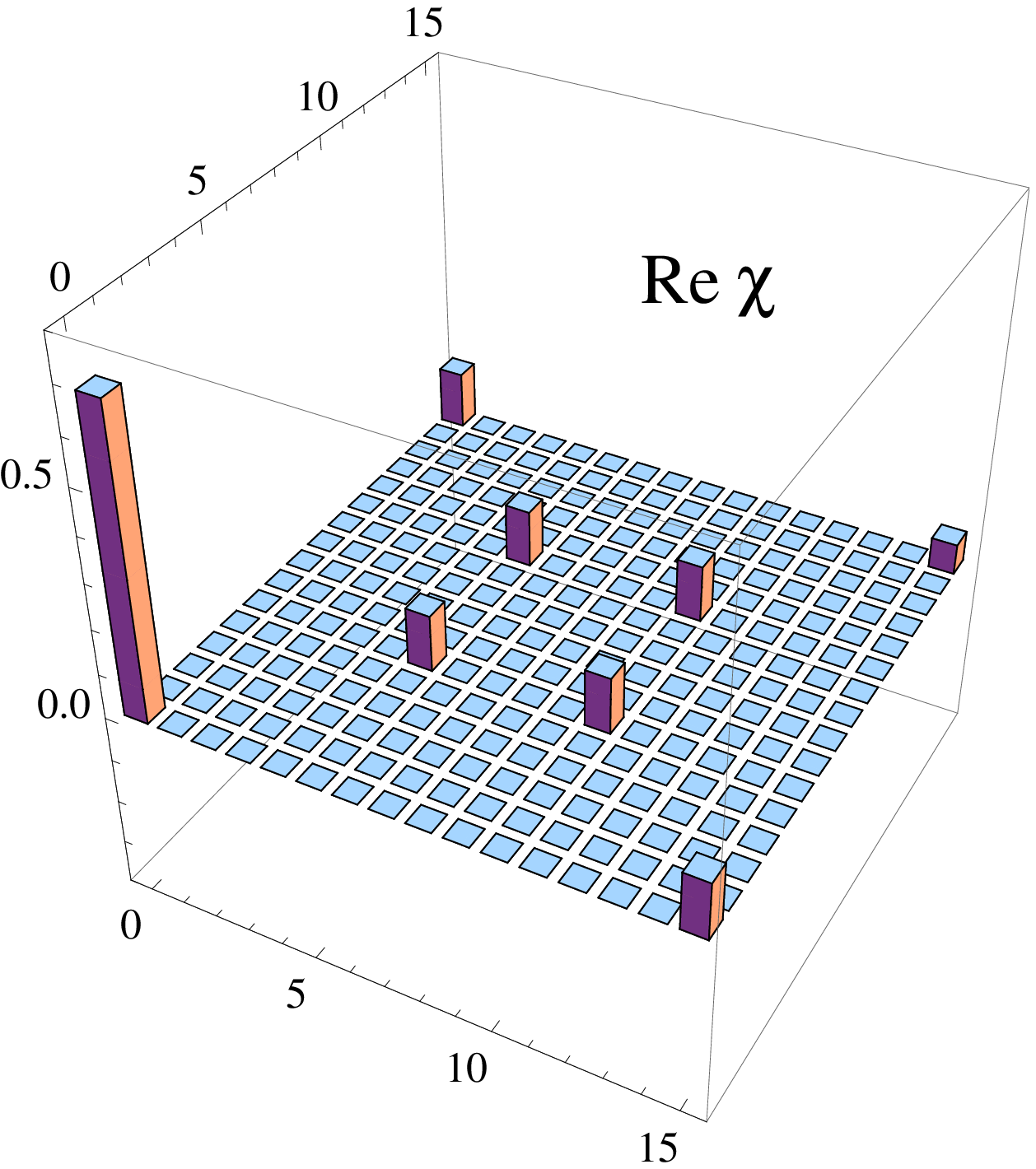}
\includegraphics[width=7.cm]{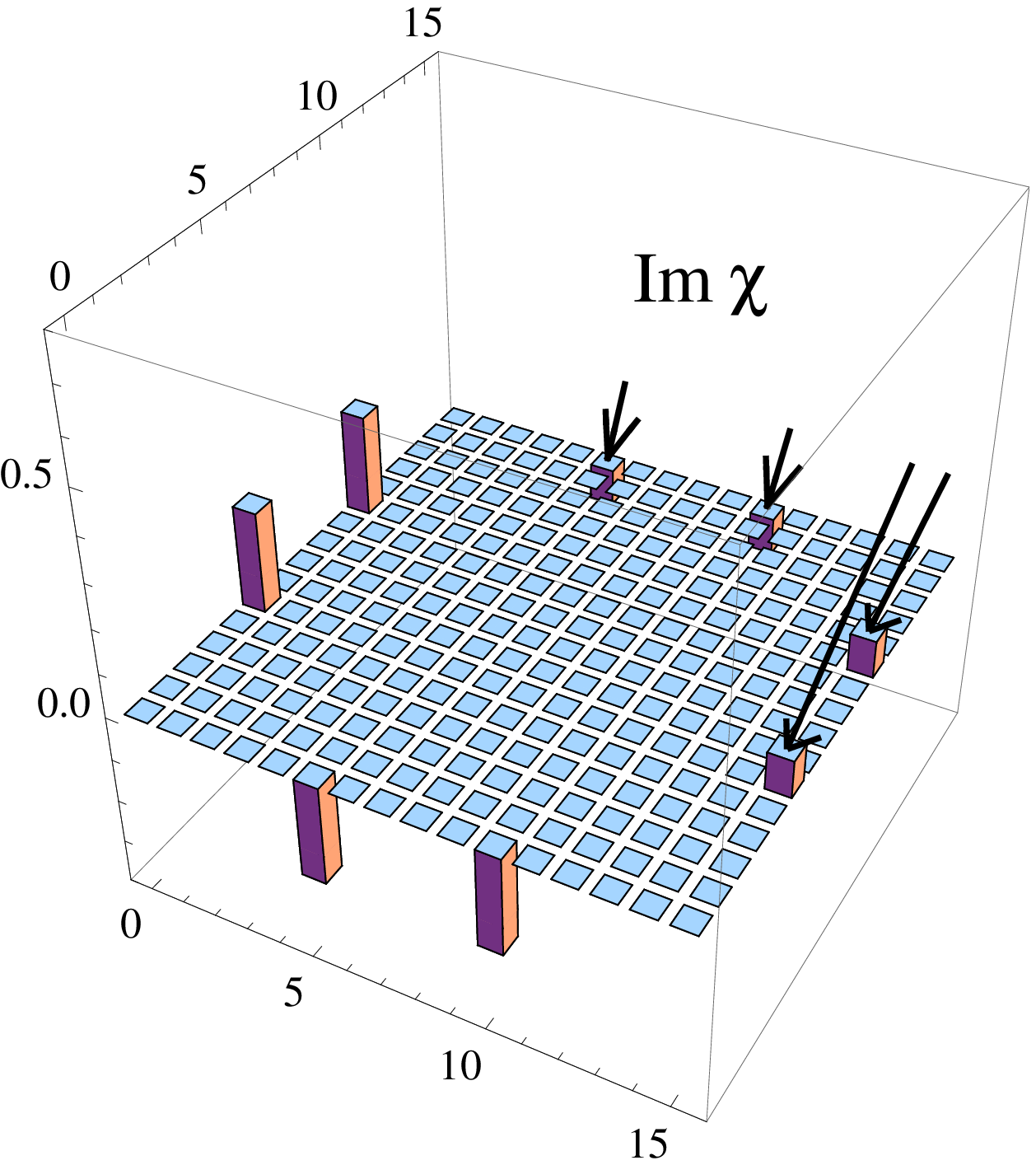}
\caption{The $\chi$-matrix (in the Pauli basis) for the $\sqrt{\rm
iSWAP}$ gate in the presence of a noisy coupling for $S/2\pi =20$
MHz and $\Gamma_s=(90$ ns)$^{-1}$. The specific feature of the noisy
coupling is the increase of the elements shown by the arrows in
comparison with their ideal value $(\sqrt{2}-1)/8$.}
 \label{f4}\end{figure}

    In contrast to the previous models, the matrix $\lambda$ for
the noisy-coupling decoherence [Eq.\ \rqn{95}] has non-zero elements
only at the positions, for which the matrix $\chi_{\rm ideal}$ is
also non-zero. As a result (not quite trivial), the noisy coupling
does not produce extra elements in the $\chi$-matrix [neither in the
first-order approximation (\ref{114}) nor in the exact solution].
The exact solution gives the following non-zero elements of $\chi$:
%TO.75.3,12
 \be
\chi_{mn}=\frac{1}{8}\left(\begin{array}{cccc}
 3+2\sqrt{2}\gamma_c&ih_+&ih_+&1\\
-ih_+&1&1&-ih_-\\
-ih_+&1&1&-ih_-\\
1&ih_-&ih_-&3-2\sqrt{2}\gamma_c
\end{array}\right),
 \e{60}
where $m,n=0,5,10,15$, $\,h_\pm=\sqrt{2}\gamma_c\pm\gamma_c^4$, and
$\gamma_c=e^{-\pi\Gamma_s/2S}$. The $\chi$-matrix for $S/2\pi = 20$
MHz and $\Gamma_s=1/(90$ ns) is shown in Fig.\ \ref{f4}.

    Despite the noisy coupling does not produce extra elements in
the $\chi$-matrix, this model also has its unique feature.
  Let us consider the imaginary elements $\chi_{5,15},\ \chi_{10,15},\
\chi_{15,5}$, and $\chi_{15,10}$, shown by the arrows in Fig.\
\ref{f4}, which all have the same absolute value $h_-/8$. From the
above formula it is easy to see that this value is larger than the
ideal value $(\sqrt{2}-1)/8\approx 0.052$ (unless decoherence is
very strong, $\Gamma_s/S>0.77$), with the maximum of 0.094 at
$\Gamma_s/S=0.22$. In all other considered models the absolute value
of these matrix elements decreases in comparison with the ideal case
[see Figs. \ref{f1}--\ref{f3} and Eq.\ \rqn{58}], so their increase
is a unique feature of the noisy coupling.

Notice that the absence of this evidence does not exclude the
possibility of noisy coupling, since the increase of the elements
marked by the arrows in Fig.\ \ref{f4} may be compensated by the
their decrease due to other decoherence mechanisms.
 Generally, the fast identification of decoherence models by their
unique features should serve only as a preliminary step, while the
accurate quantification of the decoherence mechanisms requires a
numerical best-fit procedure.

\subsection{Discussion}
\label{VIIE}

As observed and discussed above, for the considered decoherence
models the positions of the largest extra elements in the
$\chi$-matrix of the $\sqrt{\rm iSWAP}$ gate coincide with the
positions of nonzero elements of $\lambda$.
 Moreover, a comparison of Eqs.\ \rqn{61} and \rqn{62t} with
Eqs.\ \rqn{93} and \rqn{94} shows that for the largest extra
elements $\chi_{mn}\approx\lambda_{mn}\tau_g$, where $\tau_g=\pi/2S$
is the gate operation time [this statement is not correct for Eq.\
\rqn{62} in the case $|\kappa|\ll 1$, but then the elements are
small anyway]. This fact is not trivial and deserves a discussion.

    Let us consider an arbitrary two-qubit entangling gate described
by a Hamiltonian $H$ with a characteristic frequency $S$. In the
presence of a weak Markovian decoherence $\bm{L}$ the first-order
approximation \rqn{114} gives the evolution map
%TO.72.7
 \be
\bm{\mathcal L}(\tau_g) = e^{\bm{L}_{\rm coh}\tau_g}+
\int_0^{\tau_g} e^{\bm{L}_{\rm coh}(\tau_g-\tau)}\, \bm{L}\,
e^{\bm{L}_{\rm coh}\tau } \, d\tau ,
 \e{152}
where $\tau_g$ is the gate operation time. This matrix can be
transformed into $\chi$, giving the corresponding separation of
terms $\chi =\chi_{\rm ideal}+\delta \chi$.
  For a very short time so that $S\tau_g\ll 1$, the exponential
factors in Eq.\ \rqn{152} are close to one, and therefore
 \be
\delta\chi\approx\lambda\tau_g ,
 \e{154}
seemingly explaining the fact observed above. The problem, however,
is that a strongly entangling gate (such as $\sqrt{\rm iSWAP}$)
necessarily operates in the different regime, $S\tau_g\agt 1$, for
which the approximation \rqn{154} is not valid.

    We have numerically checked the relative error of the approximation
\rqn{154} for the $\sqrt{\rm iSWAP}$ gate by calculating the
dimensionless parameter $\epsilon ={\rm
Tr}|\delta\chi-\lambda\tau_g|/{\rm Tr} |\delta\chi|$, introduced
similar to Eq.\ (\ref{24}). As expected, we have obtained
$\epsilon\sim 1$ (e.g., $\epsilon =0.54$ for the energy relaxation),
confirming that the approximation \rqn{154} is invalid. However, the
inaccuracy $\epsilon$ happens to be mainly due to a large number of
small non-zero elements in $\delta\chi$, while for the largest extra
elements (where $\chi_{\rm ideal}$ is zero) the approximation
\rqn{154} unexpectedly survives. The origin of this fact is still a
puzzle for us. Nevertheless, the same useful property may hopefully
be also valid for some other quantum gates and decoherence models.

\subsection{Strongly detuned qubits}
\label{VIIF}

In Sec.\ \ref{VIIC} we have calculated the $\chi$-matrix for the
identity gate, which is realized when the two qubits are uncoupled.
However, in many experimental realizations (e.g., for capacitively
coupled phase qubits \cite{mcd05,ste06}) the qubits are permanently
coupled, and effective uncoupling is produced by strong detuning of
the qubits: $|\Delta\omega| \gg |S|$, where $\Delta \omega
=\omega_{qb,1}-\omega_{qb,2}$ is the detuning. In such experiment it
is natural to extract decoherence parameters from the
$\chi$-matrices in both situations: resonant qubits ($\sqrt{\rm
iSWAP}$) and strongly detuned qubits (that gives a more
straightforward access to decoherence parameters).
    In this subsection we analyze the $\chi$-matrix in the case of
strongly detuned qubits.

    To deal with detuned qubits we need to introduce a
rotating frame, which rotates with different frequencies for
different qubits. For this frame it is preferable to use the actual
eigenfrequencies (shifted due to the level repulsion), even though
we still use the basis of uncoupled states ($|01\rangle$,
$|10\rangle$, etc.). This produces Hamiltonian [which replaces Eq.\
(\ref{25})]
    \begin{eqnarray}
&& H=(\hbar /2) (\Delta \omega -\Delta\tilde\omega)
(|1\rangle\langle1|\otimes I- I\otimes|1\rangle\langle1|)
    \nonumber \\
&&\hspace{0.5cm}    + (\hbar S/2) (e^{-i\Delta\tilde\omega
t}|01\rangle\langle 10|+e^{i\Delta\tilde\omega t}|10\rangle\langle
01|),
    \label{H-detun}\end{eqnarray}
where $\Delta\tilde\omega = \sqrt{(\Delta\omega)^2+S^2}\times{\rm
sgn}(\Delta \omega)$ is the difference of eigenfrequencies (we
define it with the same sign as for $\Delta\omega$), and the first
term is due to the level repulsion. We assume strong detuning, which
means $|\Delta\tilde\omega|\approx |\Delta\omega|\gg |S|$; then the
first term in (\ref{H-detun}) is small since $\Delta
\tilde\omega-\Delta\omega \approx S^2/2\Delta\omega$.

    For strongly detuned qubits, Eq.\ \rqn{chi-simple-lambda} for the identity
gate $\chi$-matrix is somewhat modified, and at sufficiently small
time $t$ the $\chi$-matrix can be approximated as
    \be
\chi\approx\chi^I+\delta\chi^c+\lambda t,
    \label{chi-detuned}\ee
     where $\chi^I$ is for the
ideal identity gate, the small correction $\delta\chi^c$ comes from
the Hamiltonian \rqn{H-detun}, and the decoherence $\lambda$-matrix
is also somewhat modified (as discussed below).

    The correction $\delta\chi^c$ oscillates in time with
frequency $\Delta\tilde\omega$. After averaging over these fast
oscillations we get (in the first order in $S$) four non-zero terms:
     \be
\delta\chi^c_{09}=\delta\chi^c_{60}= -\delta\chi^c_{06}
=-\delta\chi^c_{90}=iS/4\Delta\tilde\omega.
 \e{168}
If we do not average over time, then these terms should be
multiplied by $(1-\cos \Delta\tilde\omega t)$; also there will be
four more terms with zero average:
$\delta\chi^c_{05}=\delta\chi^c_{0,10}=
-\delta\chi^c_{50}=-\delta\chi^c_{10,0}=iS \sin (\Delta\tilde\omega
t)/4\Delta\tilde\omega$. Notice that $\delta\chi^c$ is small only in
the rotating frame based on the eigenfrequencies. If, for example,
the rotating frame uses the unperturbed qubit frequencies, then both
qubits will be slowly rotating about the $z$-axis of the Bloch
sphere, that will eventually produce large terms
$\delta\chi^c_{03}$, $\delta\chi^c_{30}$, $\delta\chi^c_{0,12}$,
$\delta\chi^c_{12,0}$, $\delta\chi^c_{12,3}$, $\delta\chi^c_{3,12}$.
In an experiment the choice of the rotating frame corresponds to the
choice of the reference microwave frequencies.

    Now let us discuss contributions of the dephasing processes to
the matrix $\lambda$. It can be shown that the contribution from the
energy relaxation in qubits is still given by Eq.\ \rqn{93} (as for
the identity gate); however, the up/down rates $\Gamma_u^{(\alpha)}$
and $\Gamma_d^{(\alpha)}$ for the two qubits ($\alpha=1,2$) can now
be different from these rates for the qubits in resonance. The
difference is because these rates are proportional to the Fourier
components of the bath spectral density at the qubit frequencies
$\pm \omega_{qb,\alpha}$, and therefore changing qubit frequencies
may noticeably change the rates. For pure dephasing the contribution
to the $\lambda$-matrix is still given by Eq.\ \rqn{94} (as for the
identity gate) without any changes.

    In contrast, the contribution to the $\lambda$-matrix due to
the noisy coupling $S+s(t)$ significantly differs from Eq.\ \rqn{95}
for the identity gate. This comes from a significant change of Eq.\
\rqn{55} when we take into account the new Hamiltonian
\rqn{H-detun}.
    First, the rate $\Gamma_s=(1/4)
\int_{0}^\infty\langle s(0)s(t)\rangle dt$ should now be replaced by
$\Gamma_s'=(1/4)\int_{0}^\infty \langle
s(0)s(t)\rangle\cos(\Delta\tilde\omega t)\, dt$. This would lead to
a negligible change if the correlation time $\tau_c^{\rm NC}$ of the
noise $s(t)$ is short: $\tau_c^{\rm NC}\ll 1/|\Delta\tilde\omega|$;
however $\Gamma_s' \ll \Gamma_s$ for a long correlation time:
$\tau_c^{\rm NC}\gg 1/|\Delta\tilde\omega|$. The second change in
Eq.\ \rqn{55} is that we should delete the term $\rho_{21}$ in the
second row and similarly the term $\rho_{12}$ in the third row (this
simple change happens in the secular approximation, $\Gamma_s'\ll
|\Delta\tilde\omega|$). As a result of the changes, the contribution
to the $\lambda$-matrix due to noisy coupling has now the following
nonzero elements:
%TO.162.9
 \bea
&&\lambda_{00}=-\Gamma_s',\ \
\lambda_{0,15}=\lambda_{15,0}=\Gamma_s'/2,\nonumber\\
&&\lambda_{55}=\lambda_{10,10} =\lambda_{5,10}=\lambda_{10,5}
=\lambda_{66}=\lambda_{99}=\Gamma_s'/4,\nonumber\\
&&\lambda_{69}=\lambda_{96}=-\Gamma_s'/4.
 \ea{169}
Notice four new positions of non-zero elements in this matrix
compared to the identity gate case \rqn{95}.

   It is important to mention that positions of non-zero elements of
$\delta\chi^c$ and non-zero elements of $\lambda$ for various
decoherence models are still all different (except element
$\lambda_{00}$, which is non-zero even in $\chi^I$, and elements
$\lambda_{0,15},\lambda_{15,0}$ which appear in both correlated
dephasing and noisy coupling). Therefore, measuring $\chi$-matrix
experimentally for strongly detuned qubits gives an easy way to find
main decoherence mechanisms and quantify their parameters. Notice
that measuring $\chi$-matrix (\ref{chi-detuned}) for several times
$t$ gives a more accurate value for $\lambda$ by a simple
least-square method and also allows for checking the linearity in
time, which essentially checks that decoherence is Markovian.

\section{Conclusion}
\label{VIII}

In this paper we have discussed the effects of decoherence on the
quantum process tomography of a quantum gate. In particular, we have
introduced (Sec.\ \ref{sec-nonlocal}) dimensionless parameters,
obtainable from experimental QPT results, which characterize
nonlocality of decoherence. As an important practical example (Sec.\
\ref{VII}), we have analyzed the process matrix $\chi$ for the
two-qubit $\sqrt{\rm iSWAP}$ gate in the presence of several local
and non-local decoherence mechanisms, typical for superconducting
phase qubits. Besides presenting explicit results for the
$\chi$-matrix in the presence of decoherence (using the Pauli
basis), we have focused on finding specific patterns for each
decoherence model. These patterns may be used for a fast
identification of the most important decoherence mechanisms in an
experiment, that is an alternative to the laborious procedure of
numerical best-fitting of experimental $\chi$-matrix. Somewhat
unexpectedly, we have found that these patterns for the considered
decoherence models are to a large extent the same for the $\sqrt{\rm
iSWAP}$ and identity gates.
   In future it is interesting to study whether or not our fast-identification
approach can be applied to other quantum gates and decoherence
mechanisms.

\acknowledgments
 The work was supported by NSA and DTO under ARO grant
W911NF-08-1-0336.

\appendix

\section{Some properties of the process matrix $\chi$}
 \label{C'}

1. Let us consider the change of the matrix $\chi$ under a linear
transformation of the basis $\{E_n\}\rightarrow\{E_n'\}$. From Eq.\
\rqn{1}, using the substitution
$E_n=\sum_{n,n'=0}^{d^2-1}V_{n'n}E_{n'}'$, where $V$ is the
$d^2\!\times\! d^2$ transformation matrix (while $E_n$ are
$d\!\times\! d$ matrices), we obtain $\rho=
\sum_{m,n=0}^{d^2-1}\chi_{mn}' E_m'\rho^0E_n'^\dagger$ with
 \be
\chi'=V\chi V^\dagger.
 \e{26}
If the both bases are orthogonal, so that  ${\rm Tr}(E_n^\dagger
E_m)=Q\delta_{nm}$ and ${\rm Tr}(E_n'^\dagger E_m')=Q'\delta_{nm}$,
then $V_{nm}={\rm Tr}(E_n'^\dagger E_m)/Q'$; in this case
$\sqrt{Q'/Q}\,V$ is a unitary matrix.

2. Let us consider the transformation of the $\chi$-matrix under
unitary transformations of the initial and final states,
$\rho\rightarrow \rho'=U\rho U^\dagger$ and $\rho^0\rightarrow
\rho_0'=U_0\rho^0 U_0^\dagger$, where $U$ and $U_0$ are unitary
operators.
 From Eq.\ \rqn{1} we obtain
%TO.112.3,6
$\rho'=\sum_{m,n=0}^{d^2-1}\chi_{mn}E_m'\rho_0'E_n'^\dagger$, where
$E_n'=UE_nU_0^\dagger$, so the extra evolution of states corresponds
to the transformation of the basis $E_n$.
 If the operators $E_n$ are orthogonal, ${\rm Tr}(E_n^\dagger
E_m)=Q\delta_{nm}$, then also ${\rm Tr}(E_n'^\dagger
E_m')=Q\delta_{nm}$ and, as follows from the previous paragraph,
 \be
\rho'= \sum_{m,n=0}^{d^2-1}\chi_{mn}' E_m\rho_0'E_n^\dagger,
 \e{130}
with $\chi'$ given by Eq.\ \rqn{26}, in which $V$ is now a unitary
matrix with the elements $V_{nm}={\rm Tr}(E_n^\dagger E_m')/Q={\rm
Tr}(E_n^\dagger UE_m U_0^\dagger)/Q$.

3. Let us obtain the $\chi$-matrix for an evolution $\rho=K\rho^0
K^\dagger$ with an arbitrary linear operator $K$.
   The most important special case is the unitary evolution (then $K$ is
unitary); however, in general $K$ is an arbitrary Kraus operator
\cite{nie00}.
    Representing $K$ in the operator basis $E_n$ as
   \be
K=\sum_{n=0}^{d^2-1}k_n E_n
 \e{68a}
and comparing the evolution equation with Eq.\ \rqn{1}, we obtain
%TO.25.3,6
 \be
\chi_{mn}=k_mk_n^*.
 \e{29a}
Notice that for the orthogonal basis, ${\rm Tr}(E_n^\dagger
E_m)=Q\delta_{nm}$, the coefficients in Eq.\ (\ref{68}) can be
calculated as
    \be k_n={\rm
Tr}(E_n^\dagger K)/Q.
 \e{30}

  4. As a simple example, let us consider the process matrix $\chi^I$ for the
identity map.
 In this case
%TO.104.1,2
$ J^I_{\langle ik\rangle\langle jl\rangle}=\bm{\mathcal
L}^I_{\langle ij\rangle\langle kl\rangle}=\delta_{ik}\delta_{jl}$,
% \e{96}
and from Eq.\ \rqn{6} we obtain
%TO.104.3
 \be
\chi^I_{mn}=\sum_{i,j=0}^{d-1}({\bm E}^{-1})_{m\langle ii\rangle}
({\bm E}^{-1})_{n\langle jj\rangle}^*.
 \e{90}
For the orthogonal basis, ${\rm Tr}(E_n^\dagger E_m)=Q\delta_{nm}$,
from Eqs.\ \rqn{29a} and \rqn{30} with $K=I$, we find
 \be
\chi^I_{mn}=Q^{-2}({\rm Tr}E_m)^*{\rm Tr}E_n.
 \e{89}
This equation further simplifies when ${\rm Tr}E_n=0$ for all $n$
except for, say, $n=0$ (as in the case of the Pauli basis). Then the
basis orthogonality yields $E_0=\sqrt{Q/d} \, I$, and Eq.\ \rqn{89}
becomes
 \be
\chi^I_{mn}=(d/Q)\delta_{m0}\delta_{n0}.
 \e{88}
For the usual normalization $Q=d$, it becomes
$\chi^I_{mn}=\delta_{m0}\delta_{n0}$.

\end{document}